\PassOptionsToPackage{svgnames}{xcolor}
\documentclass[twocolumn]{aastex62}
\usepackage{graphicx}
\usepackage[flushleft]{threeparttable}
\usepackage{blindtext}
\usepackage{amsmath}
\usepackage{mathtools}
\usepackage{multirow}
\usepackage{comment}
\usepackage{booktabs}

\newcommand{\beq}{\begin{equation}}
\newcommand{\eeq}{\end{equation}}
\newcommand{\bea}{\begin{eqnarray}}
\newcommand{\eea}{\end{eqnarray}}

\begin{document}
\shortauthors{Park et al.}
\def\nar{New Astron.}

\title{Impact of Self-shielding Minihalos on the Ly$\alpha$ Forest at High Redshift}

\correspondingauthor{Hyunbae Park}
\email{hyunbae.park@lbl.gov}

\author[0000-0002-7464-7857]{Hyunbae Park}
\affil{Lawrence Berkeley National Laboratory, CA 94720, USA}

\author{Zarija Luki\'c}
\affil{Lawrence Berkeley National Laboratory, CA 94720, USA}

\author{Jean Sexton}
\affil{Lawrence Berkeley National Laboratory, CA 94720, USA}

\author{Marcelo A. Alvarez}
\affil{Lawrence Berkeley National Laboratory, CA 94720, USA}

\author[0000-0002-0410-3045]{Paul R. Shapiro}
\affil{Department of Astronomy, University Texas, Austin, TX 78712-1083, USA}

\begin{abstract} Dense gas in minihalos with masses of $10^6-10^8~M_\odot$ can shield themselves from reionization for $\sim100$ Myr after being exposed to the UV background. These self-shielded systems, often unresolved in cosmological simulations, can introduce strong absorption in quasar spectra. This paper is the first systematic study on the impact of these systems on the Ly$\alpha$ forest. We first derive the HI column density profile of photoevaporating minihalos by conducting 1D radiation-hydrodynamics simulations. We utilize these results to estimate the Ly$\alpha$ opacity from minihalos in a large-scale simulation that cannot resolve self-shielding. When the ionization rate of the background radiation is $0.03\times10^{-12}~{\rm s}^{-1}$, as expected near the end of reionization at $z\sim5.5$, we find that the incidence rate of damped Ly$\alpha$ absorbers increases by a factor of $\sim2-4$ compared to at $z=4.5$. The Ly$\alpha$ flux is, on average, suppressed by $\sim 3\%$ of its mean due to minihalos. The absorption features enhance the 1D power spectrum up to $\sim5\%$ at $k\sim0.1~h~{\rm Mpc}^{-1}~({\rm or}~10^{-3}~{\rm km}^{-1}~{\rm s})$, which is comparable to the enhancement caused by inhomogeneous reionization. The flux is particularly suppressed in the vicinity of large halos along the line-of-sight direction at separations of up to $10~h^{-1}~{\rm Mpc}$ at $r_\perp\lesssim2~h^{-1}~{\rm Mpc}$. However, these effects become much smaller for higher ionizing rates ($\gtrsim0.3\times10^{-12}~{\rm s}^{-1}$) expected in the post-reionization Universe. Our findings highlight the need to consider minihalo absorption when interpreting the Ly$\alpha$ forest at $z\gtrsim5.5$. Moreover, the sensitivity of these quantities to the ionizing background intensity can be exploited to constrain the intensity itself.
\end{abstract}

\section{Introduction}

Roughly between $10^8$ and $10^9$ yr after the Big Bang, the intergalactic medium (IGM) was reionized by UV radiation from early galaxies \citep[see reviews by][]{2001ARA&A..39...19L,2019arXiv190706653W}. During this period, the initially cold and neutral IGM transitioned to a hot plasma of $\sim$20,000 Kelvin, making it impossible for dark matter halos with masses less than $10^8~M_\odot$ to gravitationally accrete gas and form baryonic structures. Since then, $10^8~M_\odot$ has served as the minimum mass scale for the formation of large-scale structures within the IGM.

Before reionization, however, the cold neutral IGM was capable of forming structures below $10^8~M_\odot$. The first stars and galaxies are believed to have formed in minihalos (MHs) with masses $\sim 10^6~M_\odot$ at $z\sim 20 - 30$ \citep{1997ApJ...474....1T,2002Sci...295...93A,2002ApJ...564...23B,2006ApJ...652....6Y,2011ARA&A..49..373B,2013fgu..book.....L}. After these first galaxies photodissociated hydrogen molecules in the IGM \citep{1997ApJ...476..458H}, subsequent MHs could still accrete gas but were unable to form stars unless they grew massive enough ($\gtrsim 10^8~M_\odot$) to excite atomic hydrogen and enable cooling \citep[e.g.,][]{2020MNRAS.498.4887B}. Consequently, baryons within MHs that formed after the first stars likely existed as non-star-forming gas clouds. During reionization, numerous such clumps were present in the intergalactic space.

Once reionization occurs, the Jeans mass increases to $\sim10^8~M_\odot$, leading to the gradual destruction of small-scale structures due to the increased pressure of photoionized gas \citep[e.g.,][]{2016ApJ...831...86P,2020ApJ...898..149D,2021ApJ...908...96P,2023MNRAS.519.6162P,2023arXiv230504959K}. 
However, dense cores of MHs can remain neutral and serve as the sinks of the ionizing background for more than 100 Myr \citep{2004MNRAS.348..753S,2005ApJ...624..491I, 2005MNRAS.361..405I,2020ApJ...905..151N}. These self-shielded systems restrict the mean free path of the ionizing photons, impeding the growth of ionized bubbles \citep[e.g.,][]{2021ApJ...923..161N}, and influencing the large-scale morphology of reionization \citep{2006ApJ...648....1G,2009MNRAS.394..960C,2012ApJ...747..126A,2021MNRAS.504.2443B,2023MNRAS.522.2047C}. Moreover, the Ly$\alpha$ opacity of these systems can attenuate the Ly$\alpha$ emission originating from nearby star-forming galaxies \citep{2021ApJ...922..263P,2022MNRAS.512.3243S}.

The evaporation time of the MHs is highly dependent on their mass. Relatively massive MHs with a mass around $10^8~M_\odot$ retain a significant portion of their baryons even after reionization \citep{2005ApJ...624..491I,2020ApJ...905..151N}, and they are associated with Lyman limit systems in the post-reionization Universe \citep{1994ApJ...427L..13S,2003ApJ...597...66M,2010ApJ...721.1448S,2010ApJ...718..392P,2015ApJS..221....2P}, while MHs with less than $10^6~M_\odot$ lose most of their gas within $\sim 10^7$ yr. The intermediate systems, ranging from $10^6$ to $10^8~M_\odot$, undergo photoevaporate over a period of $\sim 10^8$ yr. The number density of these self-shielded systems is believed to have evolved rapidly between $z=5.5$ and $4.5$ as less massive ones evaporate earlier than the more massive ones do.

The Ly$\alpha$ forest, spectral features in spectra of distant quasars due to intervening Ly$\alpha$ absorbers, is the most effective means of probing intergalactic structures \citep[for a recent review, see][]{2016ARA&A..54..313M}.  Therefore, in this study, we will explore the potential impact of these shielded MHs on the statistical properties of the Ly$\alpha$ forest near the end of reionization. Without star formation, the MHs would follow the truncated isothermal sphere profile until they are exposed to the ionizing background. The neutral hydrogen column density of these objects can even exceed $2\times 10^{20}~{\rm cm}^{-2}$ at the cores, resulting in the damping-wing opacity of atomic hydrogen casting an extended shadow of $\gtrsim 10$ Mpc in quasar spectra. Given the high number density of the intermediate-mass ($10^6 - 10^8~M_\odot$) MHs and the challenges in subtracting damped Ly$\alpha$ absorbers (DLAs) from the high-$z$ Ly$\alpha$ forests due to low average flux, the damping-wing absorption by the self-shielded systems can significantly impact the statistics of the high-$z$ Ly$\alpha$ forest. While the absorption features of these MHs in the 21cm forest were explored by \citet{2002ApJ...579....1F}, this study is the first to focus on their impact on the Ly$\alpha$ forest.

Recently, there has been growing attention toward the high-$z$ Ly$\alpha$ forest as a promising future probe of reionization \citep{2006ARA&A..44..415F}. 
In addition to UV background fluctuations, inhomogeneous reionization induces large-scale thermal fluctuations, which leave more lasting observable signatures on the distribution and power spectrum of the Ly$\alpha$ flux even after the end of the reionization \citep[$z\gtrsim5$; e.g.,][]{2002AJ....123.2183S,2016MNRAS.463.2335N,2017ApJ...837..106O,2019MNRAS.486.4075O,2019MNRAS.490.3177W,2022MNRAS.509.6119M,2023MNRAS.521.1489M,2023MNRAS.519.6162P}. The number of quasars discovered at such high redshifts has been steadily increasing in recent yr \citep{2006AJ....132..117F,2007ApJ...662...72B,2010AJ....139..906W,2011Natur.474..616M,2013ApJ...779...24V,2015MNRAS.447.3402B,2015Natur.518..512W,2016ApJ...833..222J,2018Natur.553..473B,2020ApJ...904...26Y}, and their spectra can be utilized to constrain the details of reionization if reionization is modeled accurately \citep[e.g.,][]{2006ApJ...639L..47L,2007MNRAS.382..325B,2017ApJ...847...63O,2021MNRAS.506.2390Q}. Additionally, \cite{2018MNRAS.474.2173H} found these signals can theoretically be detected down to $z\sim2$. Thus, a substantial number of quasars to be observed at $z\lesssim 4.5$ by large-scale experiments such as the Dark Energy Spectroscopic Instrument (DESI) survey could also prove useful in this context \citep{2019MNRAS.487.1047M, 2021MNRAS.508.1262M}. 

If the absorption by MHs affects the statistics of the high-$z$ Ly$\alpha$ flux, it is crucial to consider this factor when analyzing the Ly$\alpha$ forest to constrain reionization. The amount of remaining neutral gas in MHs depends on the timing and intensity of the ionizing background. Therefore, the presence of MH absorption in the Ly$\alpha$ forest may provide extra information about reionization. In this study, we shall explore these possibilities.

Due to the significant difference in length scale between MH photoevaporation ($\sim {\rm ckpc}$) and the Ly$\alpha$ forest ($\sim 100~{\rm cMpc}$), it is computationally prohibitive to simulate the Ly$\alpha$ forest while simultaneously resolving MH self-shielding. Instead, we will employ two simulation methods at different scales to address this issue. In Section 2, we describe our 1-dimensional radiation-hydrodynamics calculation for modeling the HI column density of individual MHs after exposure to the ionizing background radiation. In Section 3, we outline our large-scale simulation for modeling the intergalactic medium and explain how we account for the small-scale absorption by MHs based on the results from Section 2. In Section 4, we present our main results on the impact of MH absorption on the Ly$\alpha$ forest. Finally, in Section 5, we summarize our work. Throughout this work, we adopt the following cosmology parameters: $h=0.675$, $\Omega_M = 0.31$, $\Omega_M = 0.31$, $\Omega_b=0.0487$, $\sigma_8 = 0.82$, and $n_s = 0.965$, which are consistent with measurements by the Planck satellite \citep{2020A&A...641A...6P}.

\begin{figure*}
  \begin{center}
    \includegraphics[scale=0.55]{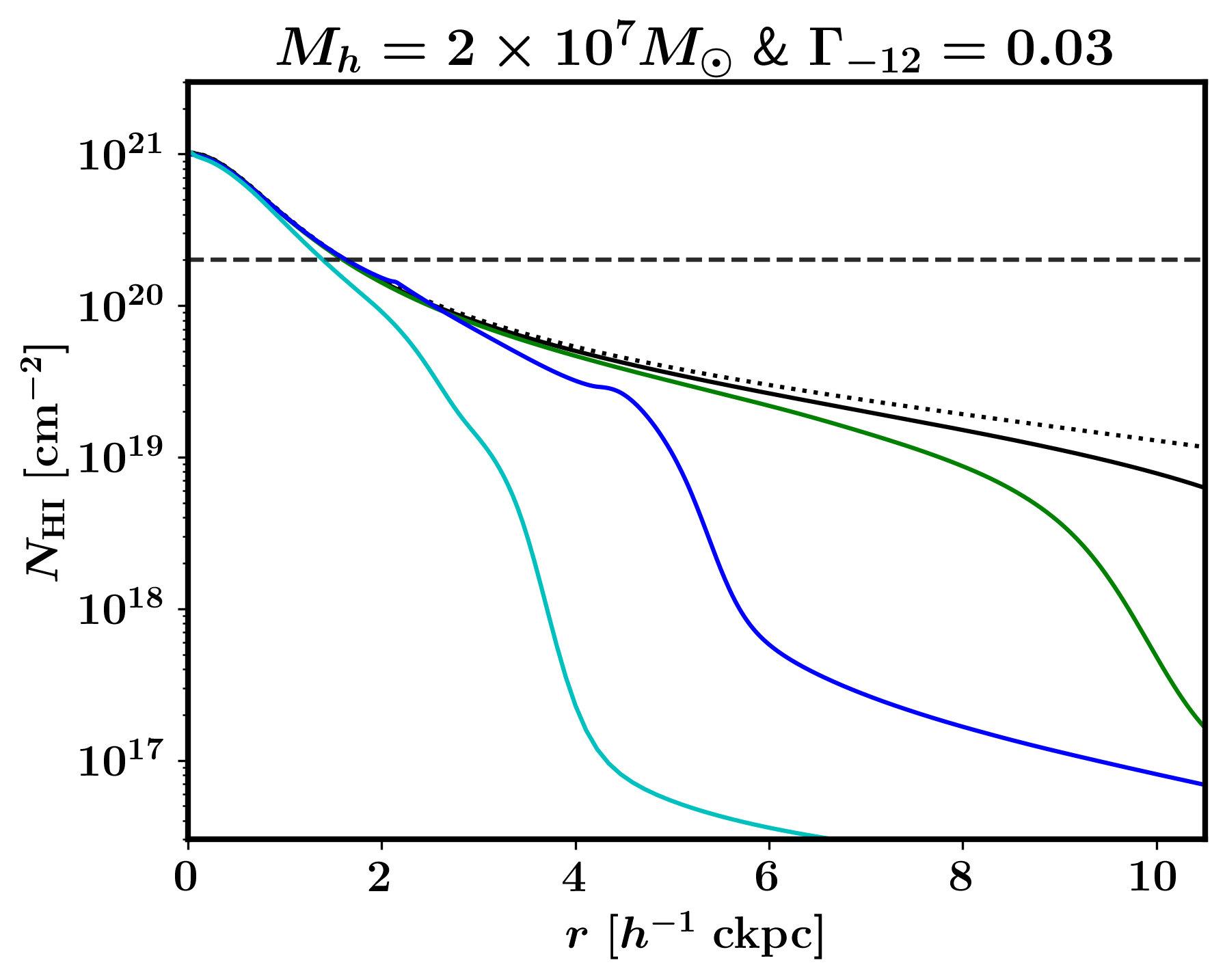}
    \includegraphics[scale=0.55]{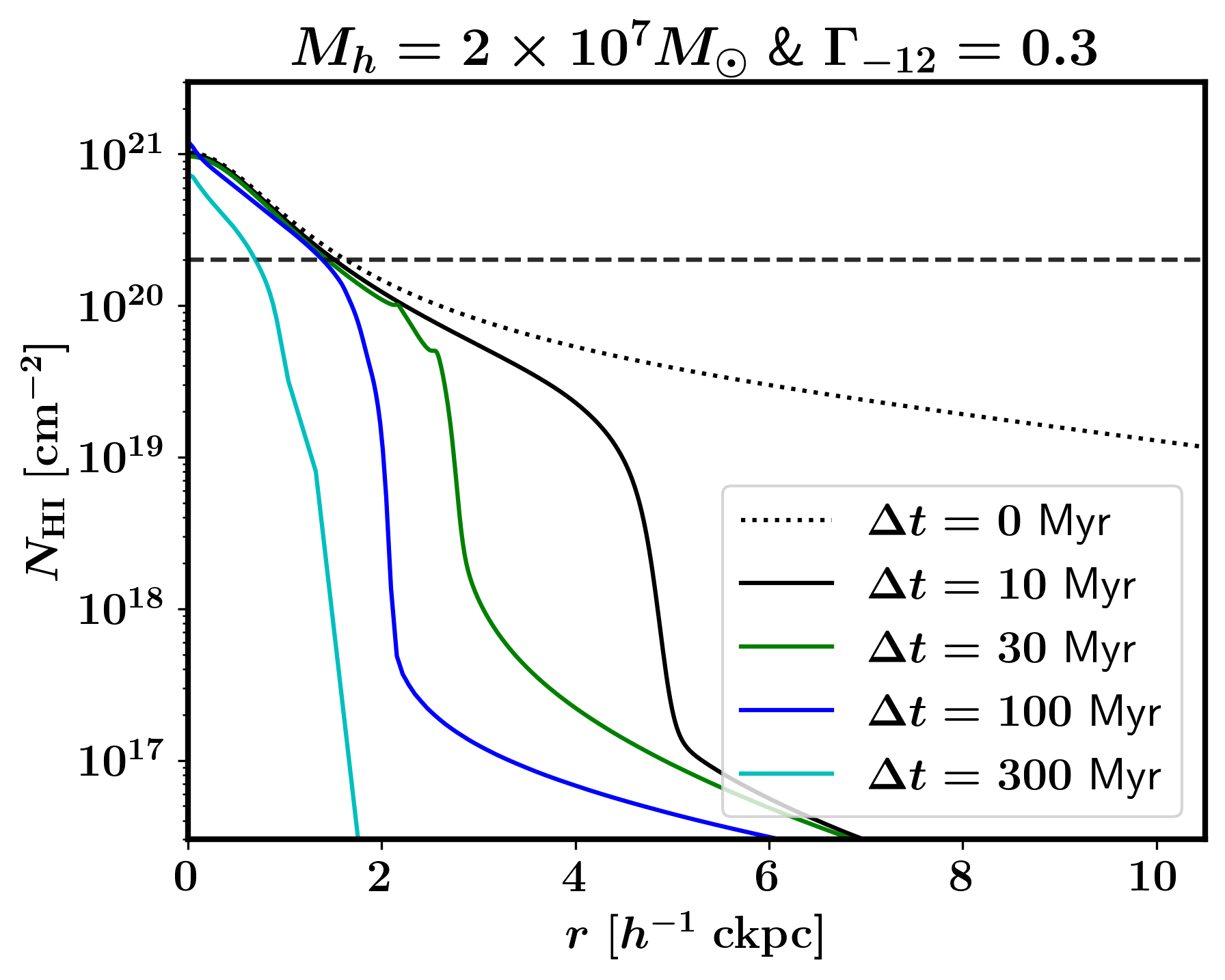}
    \includegraphics[scale=0.55]{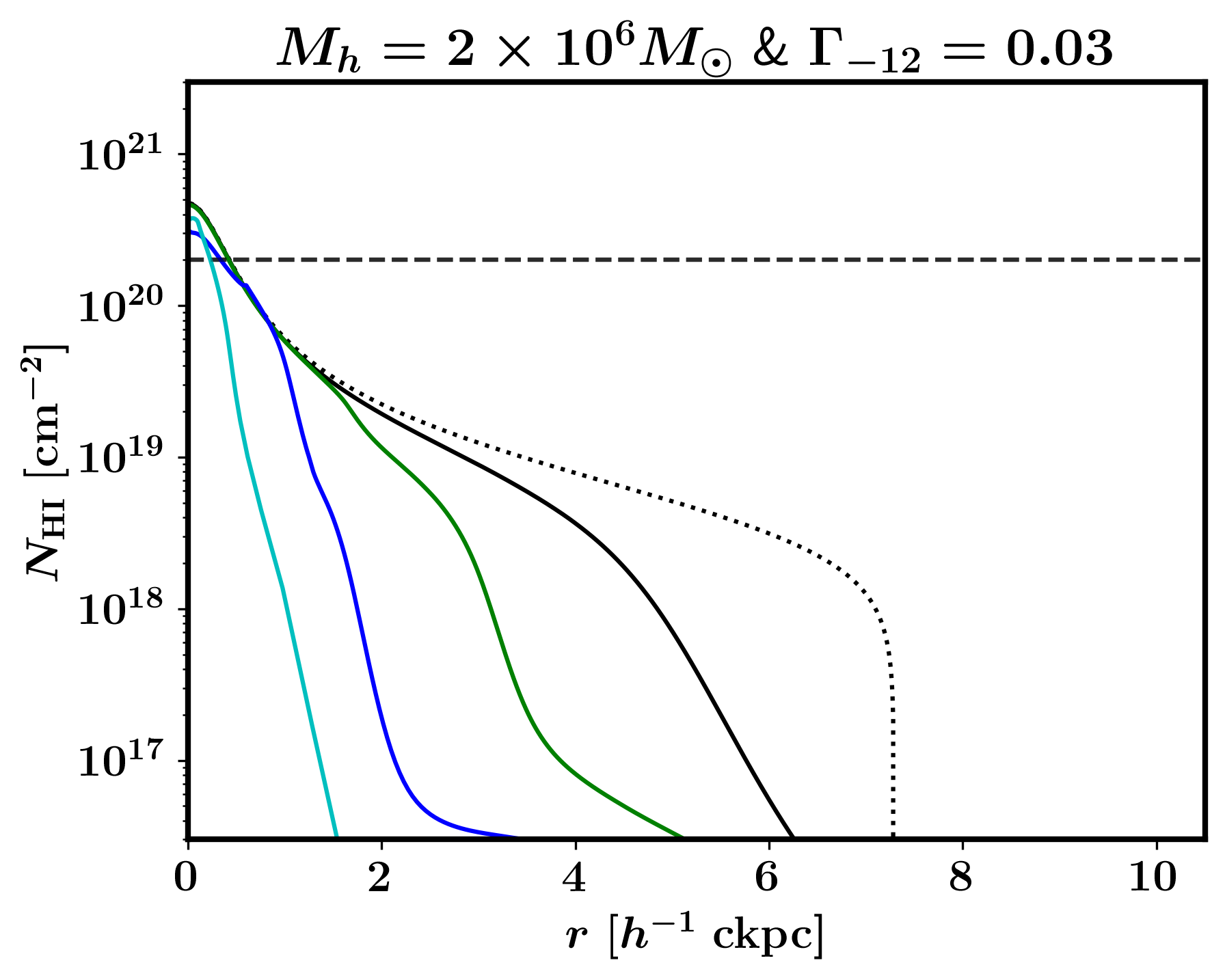}
    \includegraphics[scale=0.55]{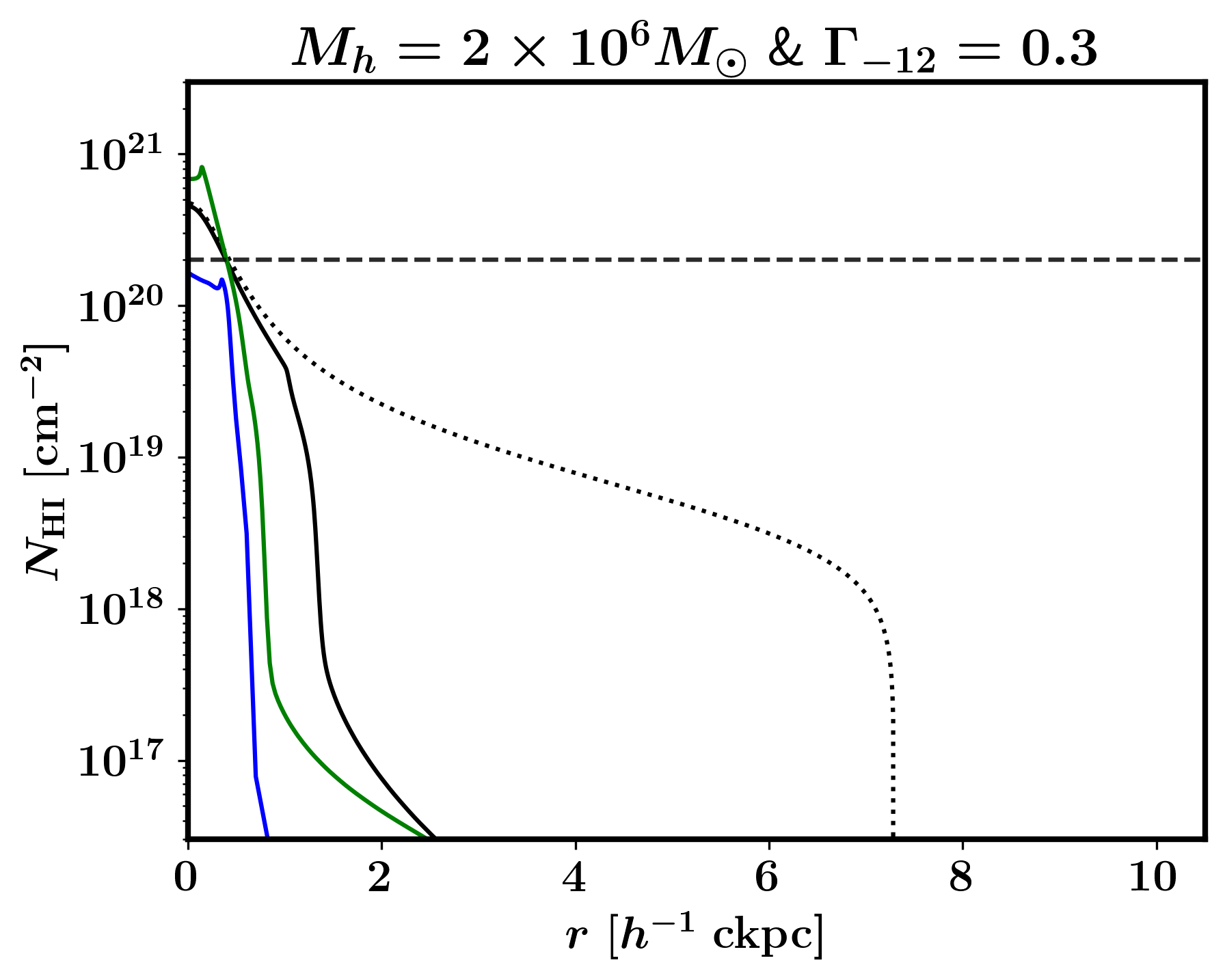}
  \caption{Radial HI column density profile from the 1D halo photoevaporation simulation. Each panel shows the time evolution of the profile with $\Delta t = 0$ (black dotted), 10 (black solid), 30 (green), 100 (blue), and 300 Myr (cyan) after exposure to the ionizing radiation, for halo masses of $M_h=2\times10^6$ (lower panels) and $2\times10^7~M_\odot$ (upper panels), as well as ionizing background intensities of $\Gamma_{12}=0.03$ (left panels) and $0.3$ (right panels).}
   \label{fig:MH_evaporation}
  \end{center}
\end{figure*}

\section{Simulation of Minihalo Photo-evaporation} \label{sec:MH_evaporation}

The process of photoevaporation in MHs involves radiative transfer and hydrodynamics at sub-kpc scales, which is generally not feasible to resolve in simulations of Ly$\alpha$ forest covering $\sim 100~{\rm Mpc}$. Therefore, we utilize the 1D radiation-hydrodynamics code, as fully described in \citet{2007MNRAS.375..881A} and further developed by \cite{2016ApJ...831...86P}, to calculate the HI column density in individual MHs exposed to the ionizing background radiation. This 1D code has been extensively tested for various cases with analytic solutions (See Appendix C of \citet{2007MNRAS.375..881A}). 

In this work, we extrapolate the fitting formula for the truncated isothermal sphere (TIS) profile to ten times the truncation radius, $r_t$, to have a reasonable description for the outskirt of the MH. We keep track of $N_{\rm sh}=$ 10,000 radial shells linearly spaced from $r=10^{-3}~r_t$ to $10~r_t$.
We bound the outermost shell with the pressure of that shell at the initial time step, which becomes practically negligible as soon as the ionization of outer shells photoheats the gas above 10,000 K. This setup is identical to that in the appendix of \cite{2016ApJ...831...86P}. 

For the initial conditions, we adopt the TIS profile, which represents the equilibrium state of MHs in the absence of star formation \citep{1999MNRAS.307..203S}. The fitting formula for this TIS profile is provided in Appendix A of \cite{1999MNRAS.307..203S}. To describe the outskirts of MHs, we extend this TIS profile to 10 times the truncation radius $r_t$ using the power-law slope at $r_t$. The mass inside $r_t$ is similar to the virial mass. In this calculation, we consider $2,~3,~5,~7,~10,~20,~30,$ and $50$ million $M_\odot$ to cover a mass range that would survive the ionizing background for a significant amount of time ($\gtrsim 10^8$ yr). The code tracks 10,000 radial shells between $10^{-3}$ and $10~r_t$. While the profile shape depends on the redshift of collapse, $z_{\rm col}$, previous works have shown that the photoevaporation rate is relatively insensitive to this quantity \citep{2004MNRAS.348..753S,2020ApJ...905..151N}.

For the ultraviolet background (UVB) radiation, we adopt the blackbody spectrum of $T_{\rm bb}=$ 100,000 K. We truncate the UVB spectrum above 54.4 eV to account for absorption by HeII. The radiation is thought to contain starlight from O- and B-type stars with a temperature range from 30,000 to 50,000 K, as well as EUV and X-ray radiation from accreting neutron stars and quasars \citep[e.g.,][]{2012ApJ...746..125H}. However, the exact spectral shape of the radiation remains unknown. The photoevaporation of MHs depends on the hardness of the ionizing radiation background, as higher-energy photons result in a thicker ionizing layer on MHs due to their longer mean free path \citep{2005MNRAS.361..405I,2020ApJ...905..151N}. The hardness of the spectrum we adopted falls between the hardness of these two components. 

For simplicity, we assume a constant and isotropic ionizing radiation with $J_{-21}=0.01$ and $0.1$. Here, $J_{-21}$ represents the angle-averaged intensity, $\int I_\nu d\Omega/(4\pi)$, at the Lyman limit in units of $10^{-21}~{\rm erg}~{\rm cm}^{-2}~{\rm s}^{-1}~{\rm Hz}^{-1}~{\rm sr}^{-1}$. Another commonly used quantity in the literature is 
\bea
\Gamma_{-12} \equiv [10^{12}~{\rm s}]\int^\infty_{\rm 13.6~eV}d\nu \frac{\sigma_\nu}{h\nu} \int I_\nu d\Omega,
\eea
where $\sigma_\nu$ is the cross section of hydrogen atoms to ionizing photons. This quantity represents the ionization rate per H atom in units of $10^{-12}~{\rm s}^{-1}$. Our choices of ionizing intensity, $J_{-21}=0.01$ and $0.1$, correspond to $\Gamma_{-12}= 0.03$ and $0.3$, respectively. We will refer to this value when referring to each case in this paper. 

Measurements from the quasar proximity zone suggest a steep increase of $\Gamma_{-12}$ near the end of reionization, with the value rising to $\Gamma_{-12}\sim1$ from a lower, yet unknown, value (see \cite{2023MNRAS.525.4093G} for the latest measurement). As a typical value for reionization, we will consider $\Gamma_{-12}=0.03$ based on the reionization model by \citet[][see their Fig. 4]{2021MNRAS.507.6108O}.

\begin{figure}
  \begin{center}
    \includegraphics[scale=0.55]{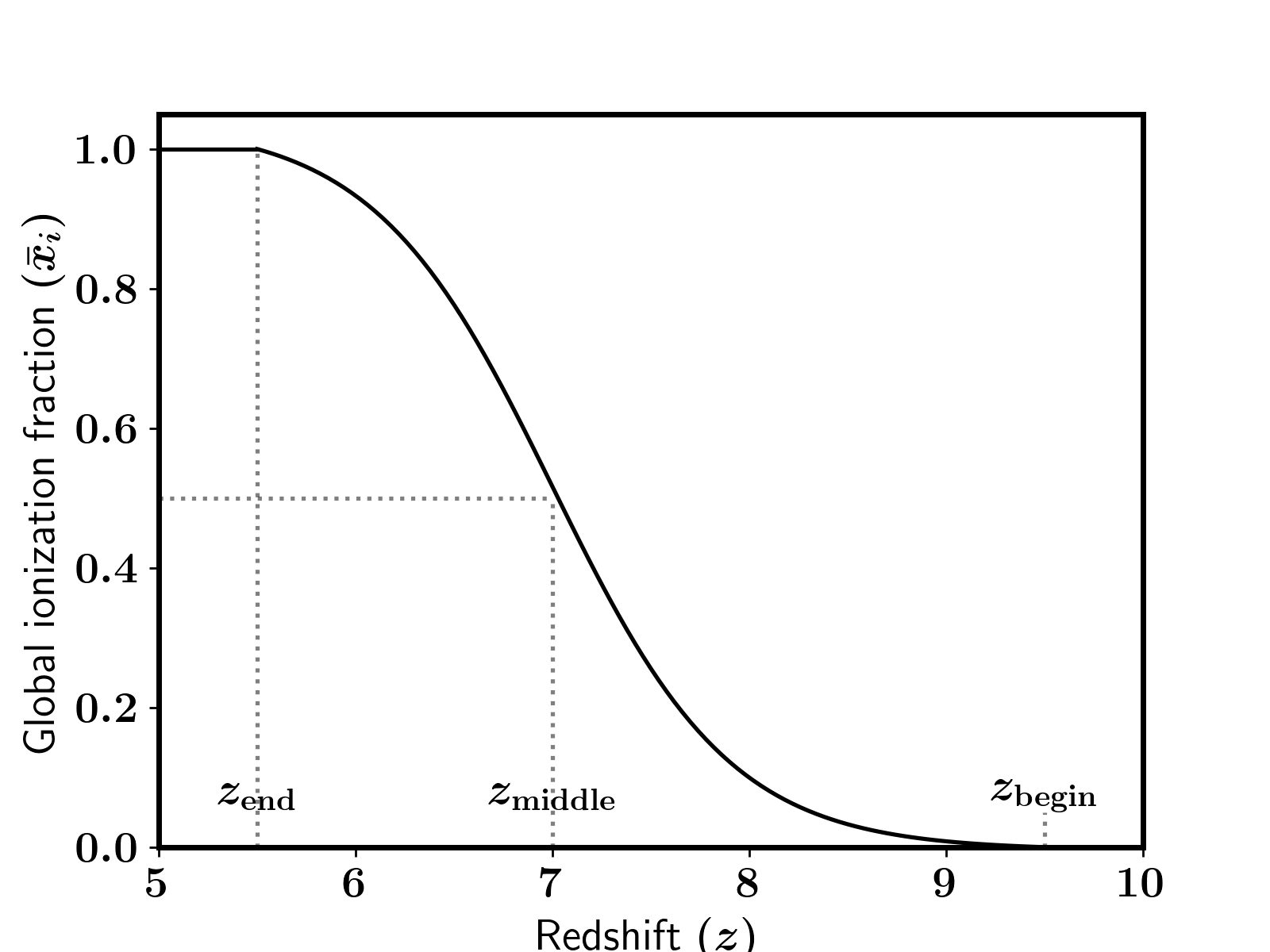}
  \caption{Volume-averaged global reionization history of the inhomogeneous reionization model used in this work. }
   \label{fig:Xfrac}
  \end{center}
\end{figure}

Figure~\ref{fig:MH_evaporation} shows the time evolution of the radial HI column density profile for halo masses of $2\times 10^6$ and $2\times10^7~M_\odot$ and ionizing intensity of $\Gamma_{-12} = 0.03$ and $0.3$. In the initial conditions given by the TIS profile, the HI column density exceeds $2\times 10^{20}~{\rm cm}^{-2}$ at the core for both masses. When exposed to the lower ionizing intensity $\Gamma_{-12}=0.03$ of the reionization era, the HI column density remains nearly unchanged for both halo masses even after 300 Myr. However, with the higher intensity $\Gamma_{-12}=0.3$ of the post-reionization era, the core of the $2\times10^7~M_\odot$ halo remains as a DLA even after 300 Myr, while the core of the $2\times10^6~M_\odot$ halo loses most of its HI gas after 100 Myr and no longer appear as a DLA. 

The HI column density profiles show that MHs with several million solar masses can withstand the ionizing radiation during reionization and manifest as DLAs for $\sim 100$ Myr after reionization. More massive MHs above $10^7~M_\odot$ would survive the stronger ionizing background of the post-reionization Universe and remain as DLAs for much longer durations. These findings align with the results of \cite{2004MNRAS.348..753S} and \cite{2020ApJ...905..151N} using similar simulations. In Section~\ref{sec:Lya-forest}, we will utilize these HI density profiles to estimate the HI column density in FoF halos of the large-scale cosmological simulation.

\begin{figure*}
  \begin{center}
    \includegraphics[scale=0.65]{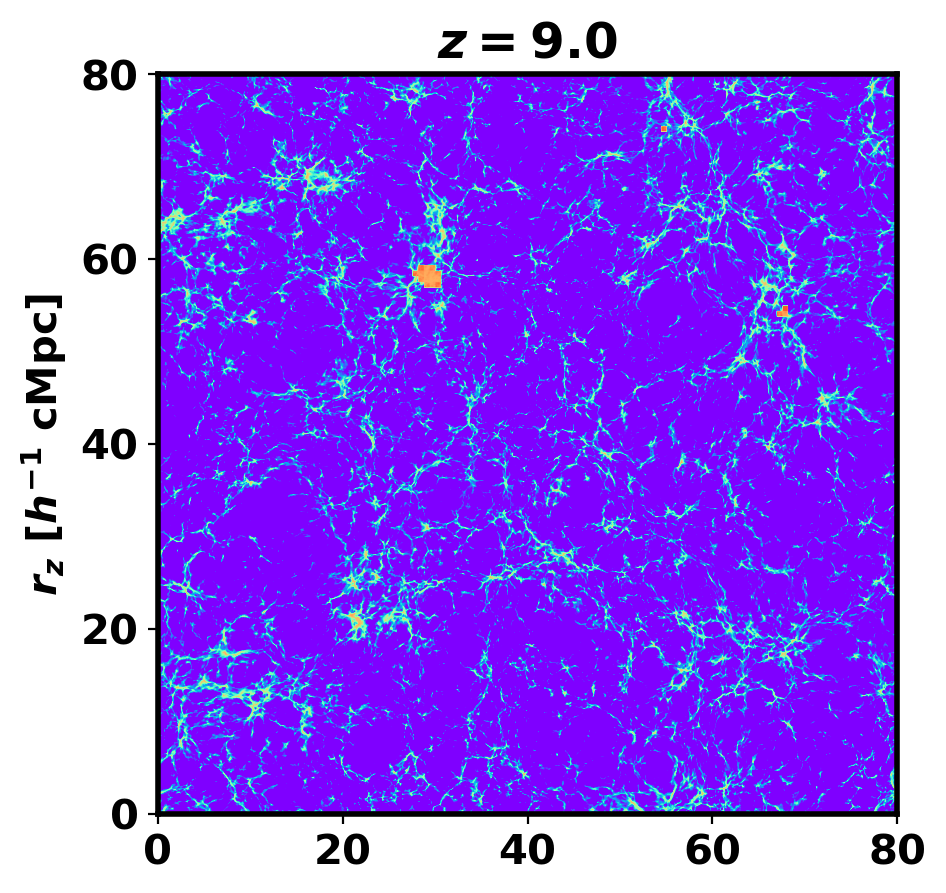}
    \includegraphics[scale=0.65]{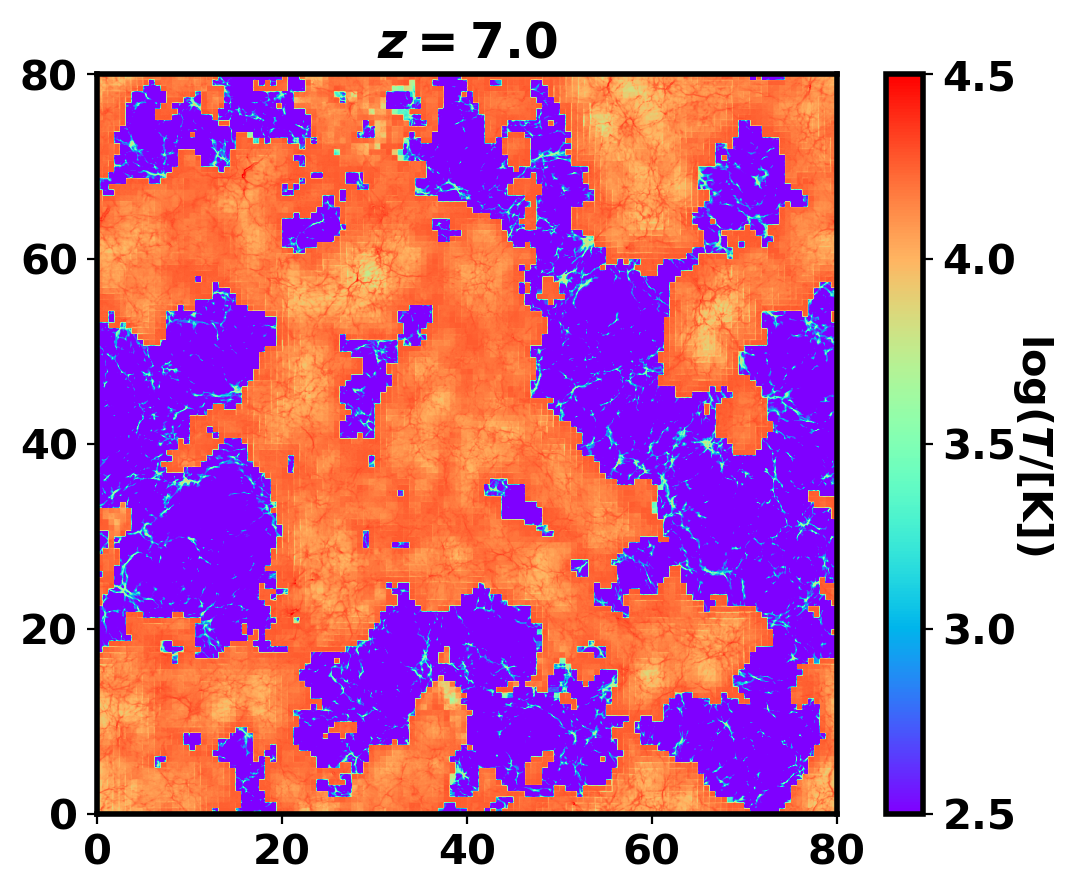}
    \includegraphics[scale=0.65]{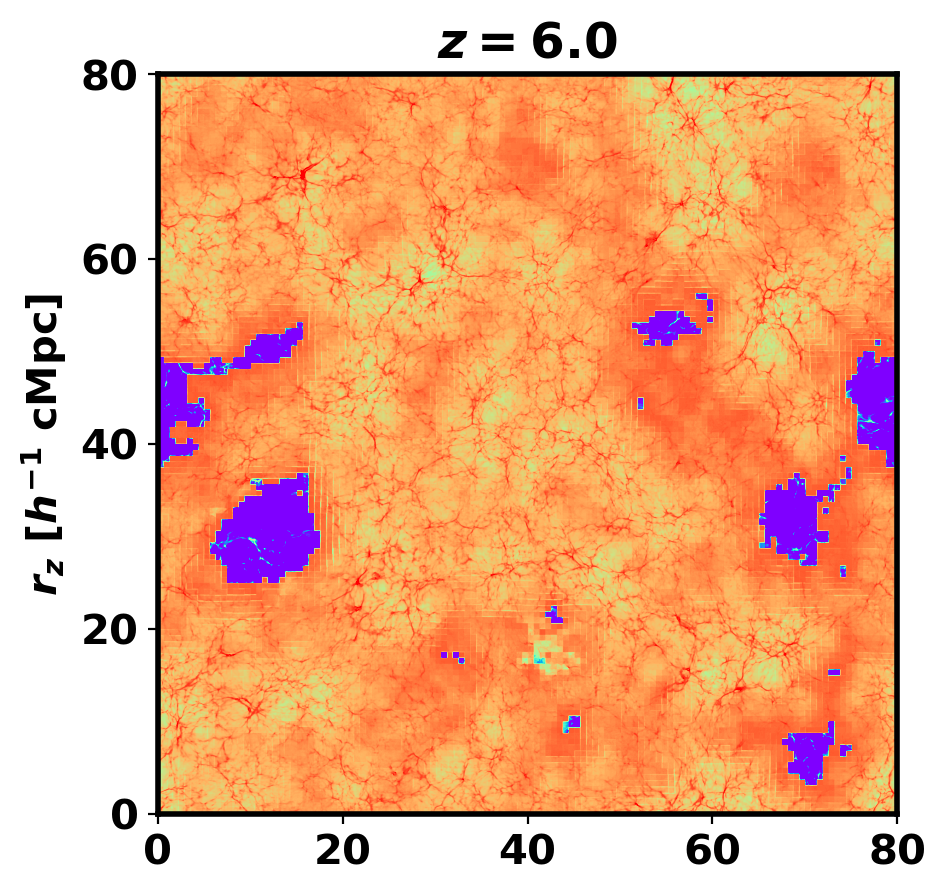}
    \includegraphics[scale=0.65]{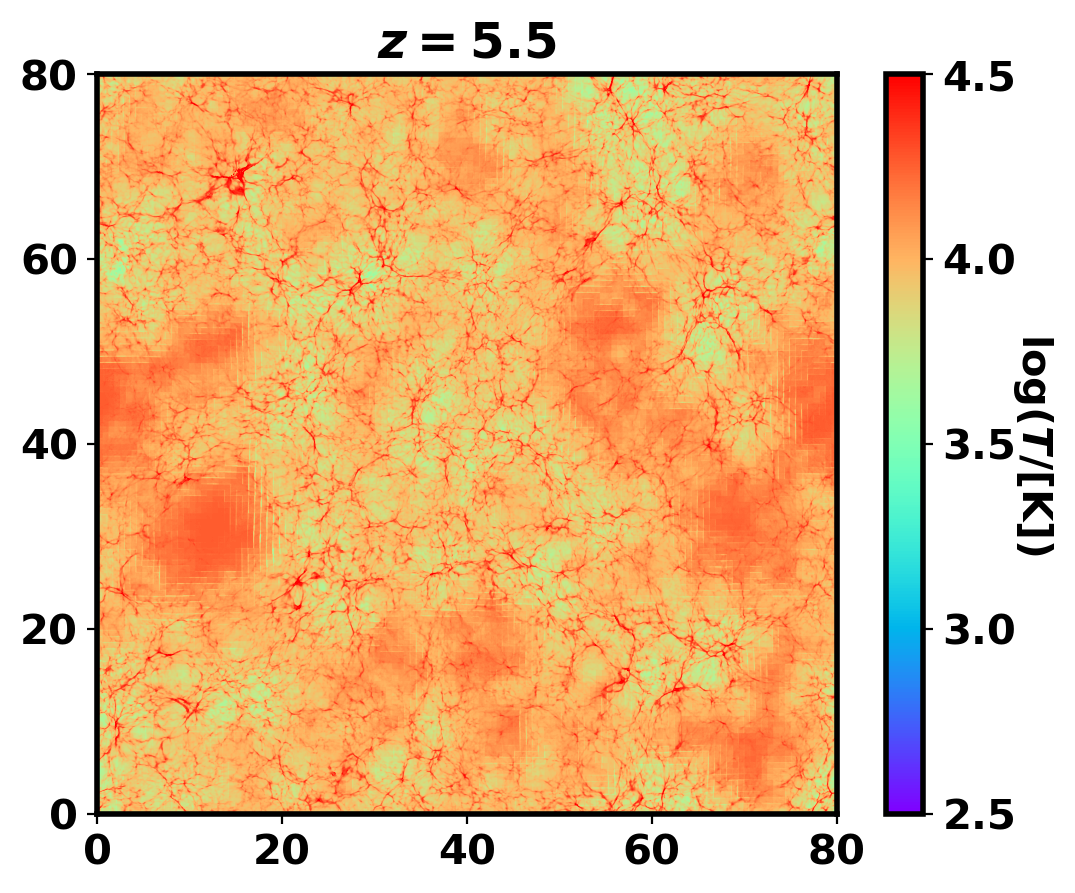}
    \includegraphics[scale=0.65]{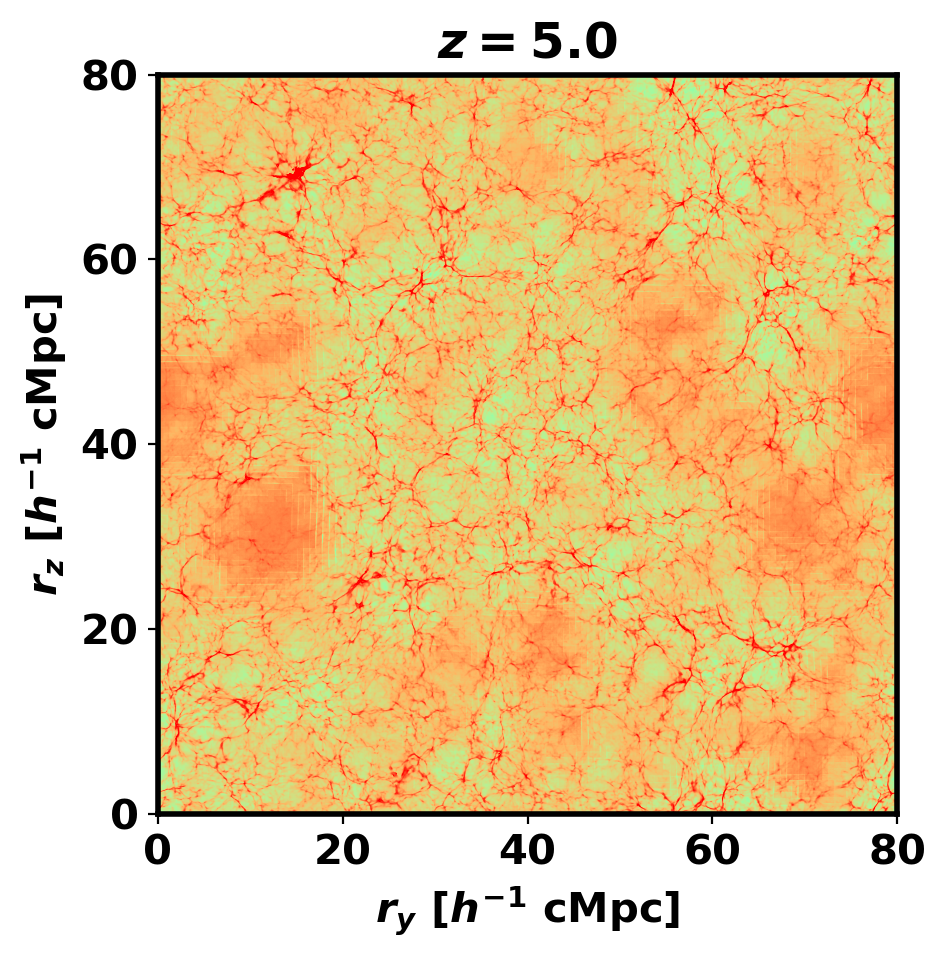}
    \includegraphics[scale=0.65]{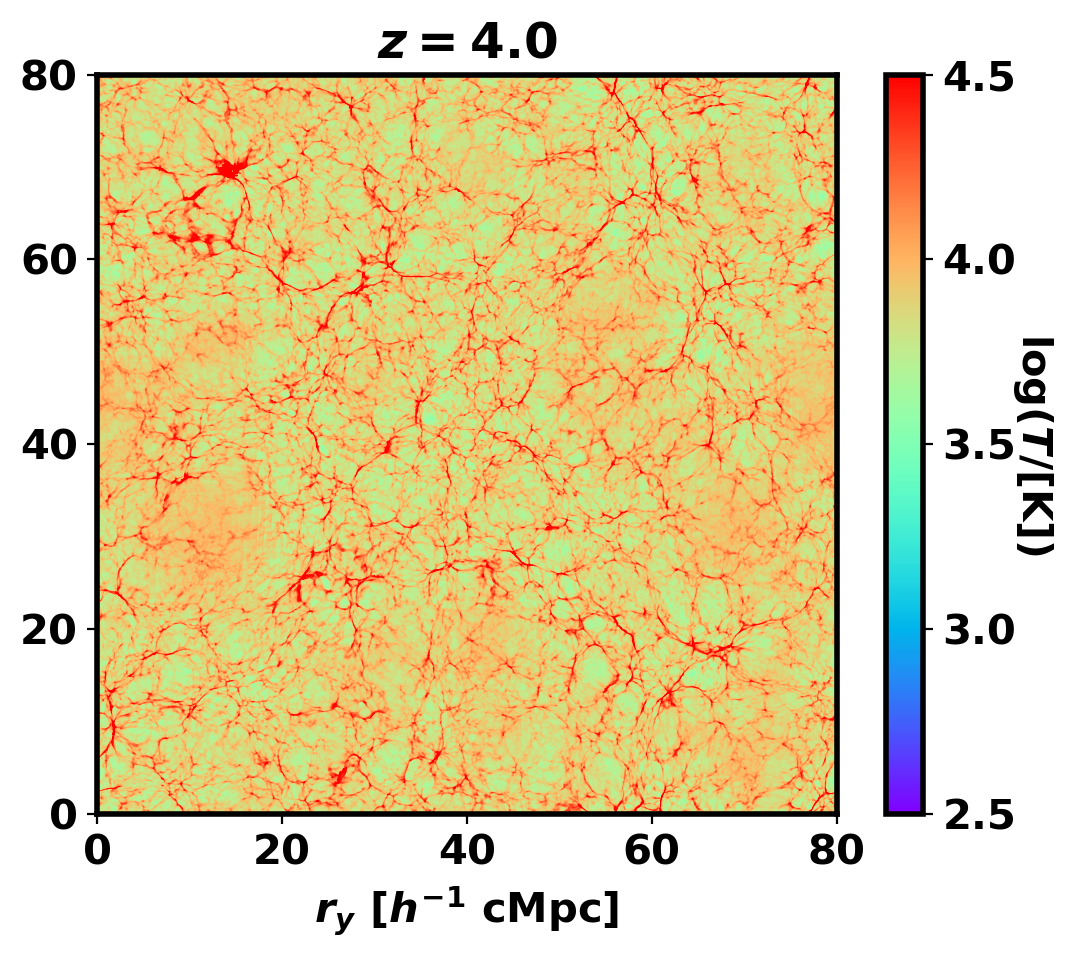}
  \caption{Temperature map of a slice from the Nyx simulation with the inhomogeneous reionization model. The six panels depict the snapshots at $z=9,~7,~6,~5.5,~5,$ and $4$ with the color representing the temperature. Reionization takes place between $z=9$ and $5.5$ in the model used in this work.}
   \label{fig:Tmap}
  \end{center}
\end{figure*}

\section{Simulation of Ly$\alpha$ Forest} \label{sec:Lya-forest}

We employ Nyx as our simulation tool for modeling the evolution of the intergalactic medium. Nyx is a cosmological $N$-body/hydrodynamics code specifically designed for efficient execution on massively parallel computers \citep{2013ApJ...765...39A}. It solves the hydrodynamic equations for gas using a grid-based approach, while dark matter is represented by collisionless particles. Nyx exclusively relies on the particle-mesh calculation for gravitational dynamics. This allows Nyx to focus on simulating mildly overdense structures that are relevant to the Ly$\alpha$ forest while saving computational time by excluding particle-particle gravity calculations for highly overdense structures where Ly$\alpha$ transmissivity is negligible. 

To ensure convergence in the Ly$\alpha$ flux statistics, it is necessary to simulate a volume of $\gtrsim 100$ Mpc while resolving structures down to scales of $\sim 10$ kpc \citep{2015MNRAS.446.3697L}. We initialize the matter density field within an $80~h^{-1}~{\rm Mpc}$ box at $z_i=199$ using $4096^3$ dark matter particles with $6.5\times 10^5~h^{-1}~~M_\odot$ each. Once the simulation starts, the gas density and velocity are initialized on a $4096^3$ mesh, assuming baryons trace dark matter.

\subsection{Inhomogeneous Reionization}

To model inhomogeneous reionization, we use a hybrid scheme that implements reionization physics into large-scale cosmological hydrodynamics simulations based on a small set of phenomenological input parameters (see \citealt{2017ApJ...837..106O}). During the onset of reionization, each cell is photoheated by $\Delta T = 20,000$ K according to the input reionization redshift ($z_{\rm re}$) field, following \citet{2019MNRAS.486.4075O}. After the cell has been exposed to reionization, the UVB model of \citet[][``Middle HI Reionization'']{2017ApJ...837..106O} is assumed to calculate the temperature evolution.

We calculate $z_{\rm re}$ on a $128^3$ grid according to the method presented by \cite{2013ApJ...776...81B}, where $z_{\rm re}$ is calculated by smoothing and rescaling the initial density field with a Gaussian filter of $1~h^{-1}~{\rm Mpc}$. Following the approach of \cite{2022ApJ...927..186T}, we rescale the smoothed density field to match a desired global reionization history, $\bar{x}_e(z)$, while preserving the order among the cell values (i.e., higher $z_{\rm re}$ for higher smoothed density). In this study, we adopt a simple analytic model with free parameters representing the beginning ($z_{\rm begin}$), middle ($z_{\rm middle}$), and end of reionization ($z_{\rm end}$). The model is given by
\bea
\bar{x}_e(z) = 0.5\frac{\tanh[A(z-z_{\rm middle})] + 1}{X_e - X_b} - X_b,
\eea
where $X_b = 0.5(\tanh[A(z-z_{\rm begin})]+1)$, $X_e = 0.5(\tanh[A(z-z_{\rm end })]+1)$, and $A=1.1$. The reionization history from this toy model is shown in Figure~\ref{fig:Xfrac}. We choose $z_{\rm begin}=9.5$, $z_{\rm middle} = 7$, and $z_{\rm end}=5.5$ in this work. The choice of $z_{\rm end}=5.5$ is consistent with the recent measurement of the Lyman limit mean free path at $z\ge 5.5$, suggesting that reionization ended at $z\lesssim6$ \citep{2021MNRAS.508.1853B}. In future applications, these parameters will be varied to explore different reionization histories and study the dependence of the Ly$\alpha$ forest on these parameters. Also, the input reionization model can easily be replaced with a more sophisticated one such as 21CMFAST \citep{2011MNRAS.411..955M}. To assess the impact of inhomogeneous reionization, we also conducted another simulation with the same initial conditions but with $z_{\rm re}=7$ assigned everywhere in the simulation volume. 

The gas temperature maps in Figure~\ref{fig:Tmap} provide an overview of the simulation and impact of the inhomogeneous reionization at large scales\footnote{We denote the $x$, $y$, and $z$ coordinates in our simulations as $r_x$, $r_y$, and $r_z$, respectively, to avoid confusion between the $z$ coordinate and the redshift $z$.}. Between $z=9$ and 5.5, reionization heats the IGM from $100$ K or below to 20,000 K, starting from overdense regions and progressing toward underdense regions. By $z=5.5$, reionization is complete, but there is still a mild large-scale variation in temperature due to regions that reionized later having less time to adiabatically cool. This reionization relic is visible at $z=5$, but it becomes mostly diluted by $z=4$.

\subsection{Ly$\alpha$ Opacity of MHs} \label{sec:method}

\begin{figure}
  \begin{center}
    \includegraphics[scale=0.55]{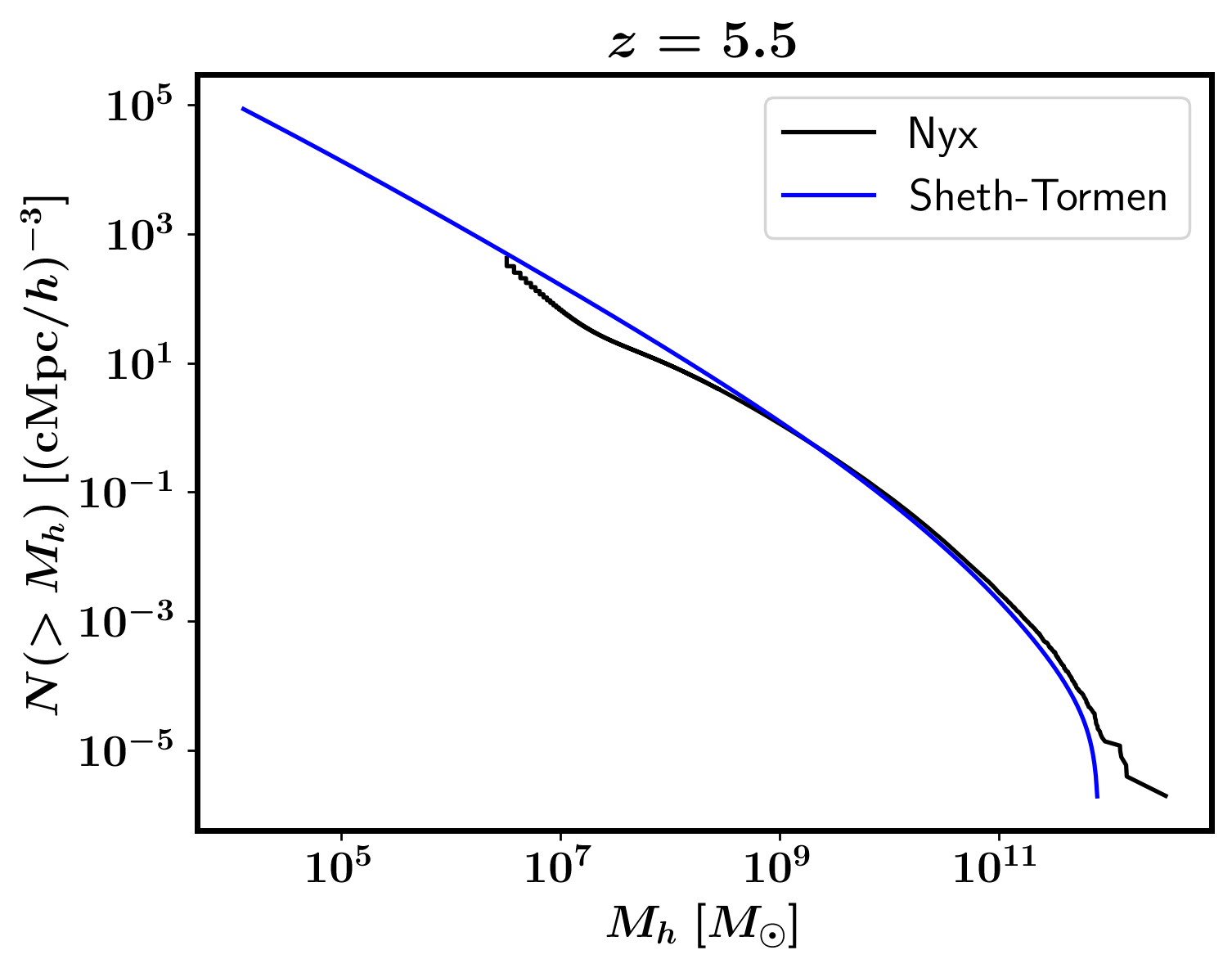}
  \caption{Cumulative halo mass function from the $z=5.5$ snapshot of our Nyx simulation run (black) compared to that from the Sheth-Tormen function (blue).}
   \label{fig:MF}
  \end{center}
\end{figure}

\begin{figure*}
  \begin{center}
    \includegraphics[scale=0.5]{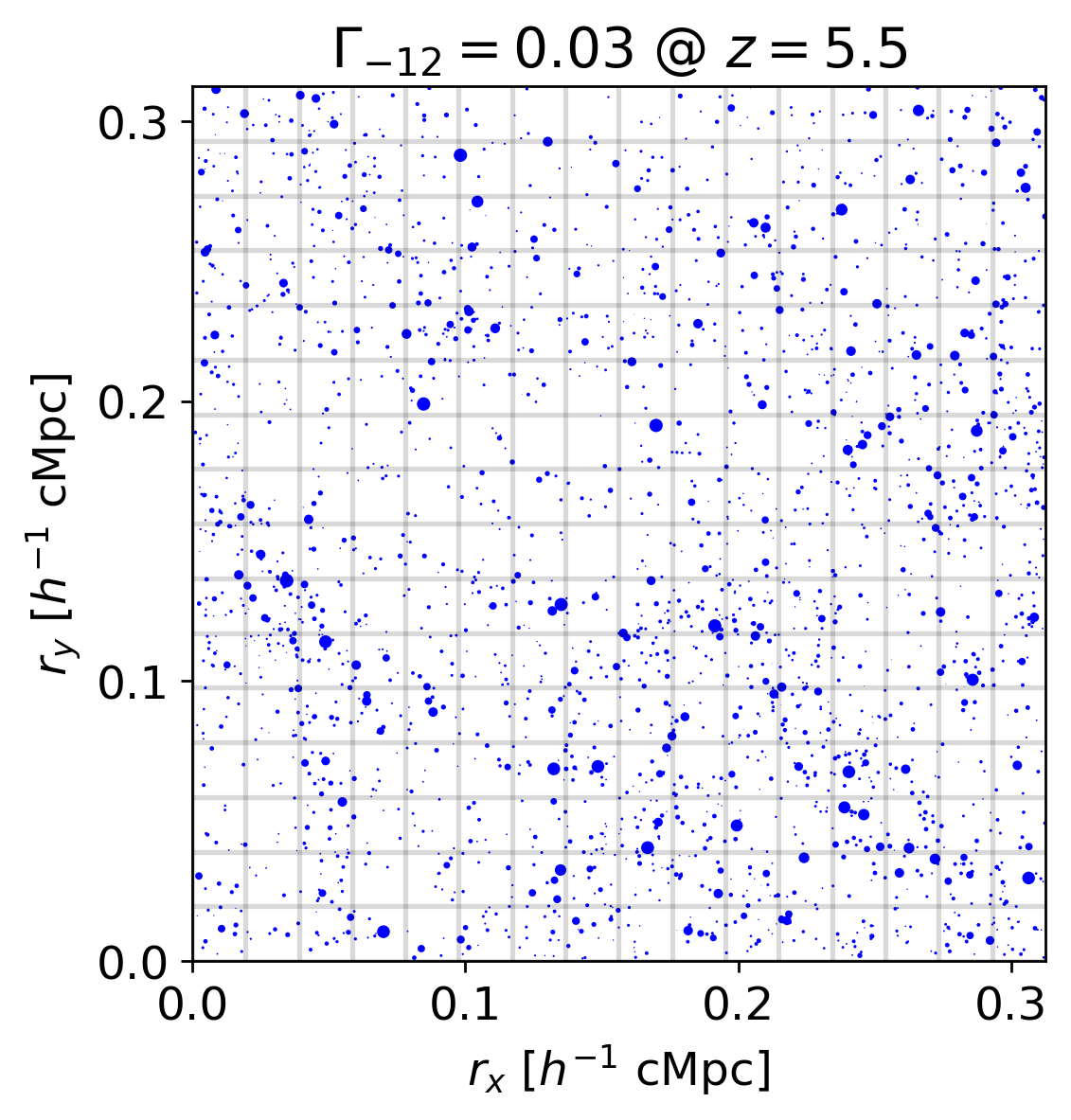}
    \includegraphics[scale=0.5]{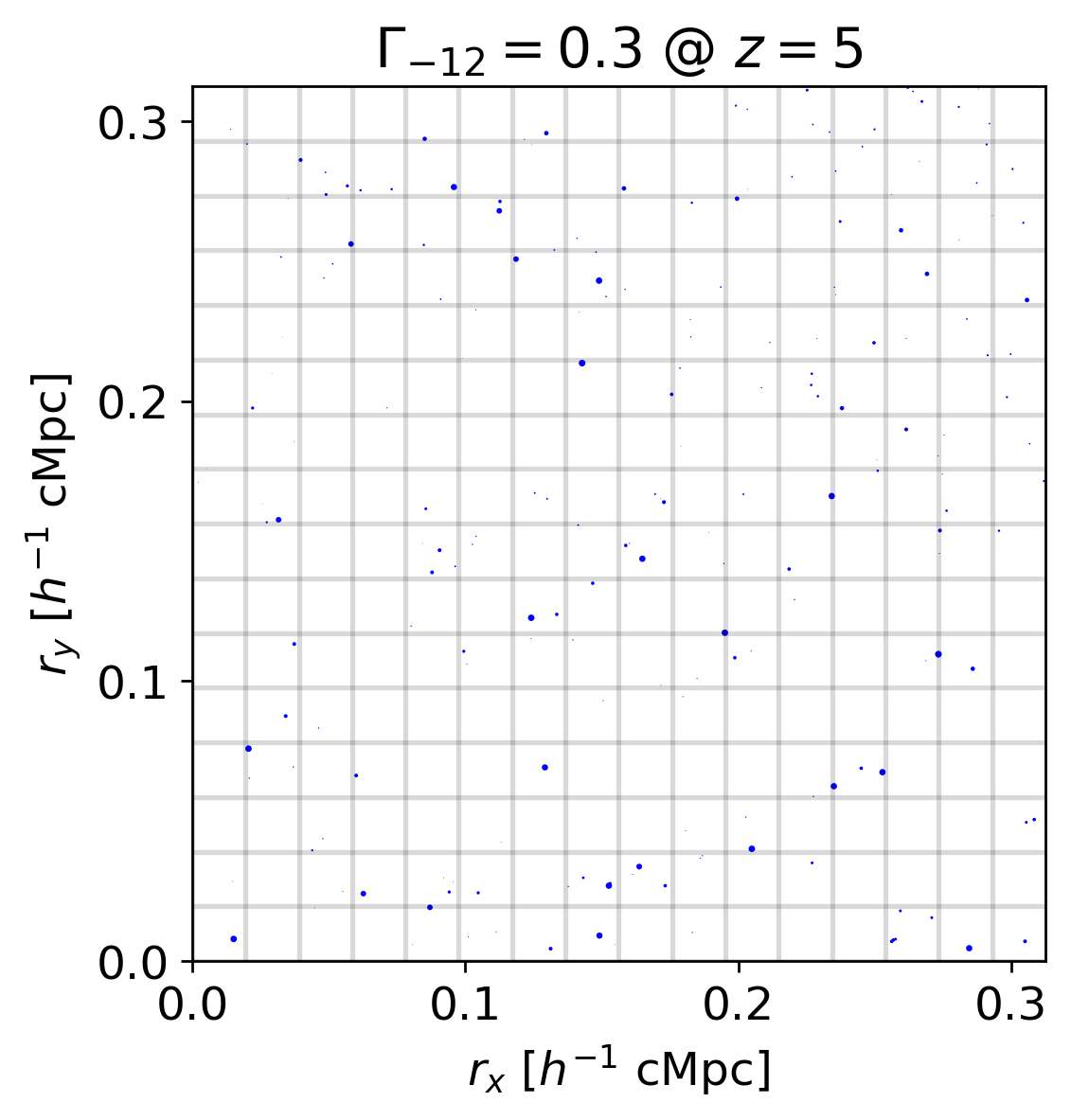}
    \includegraphics[scale=0.5]{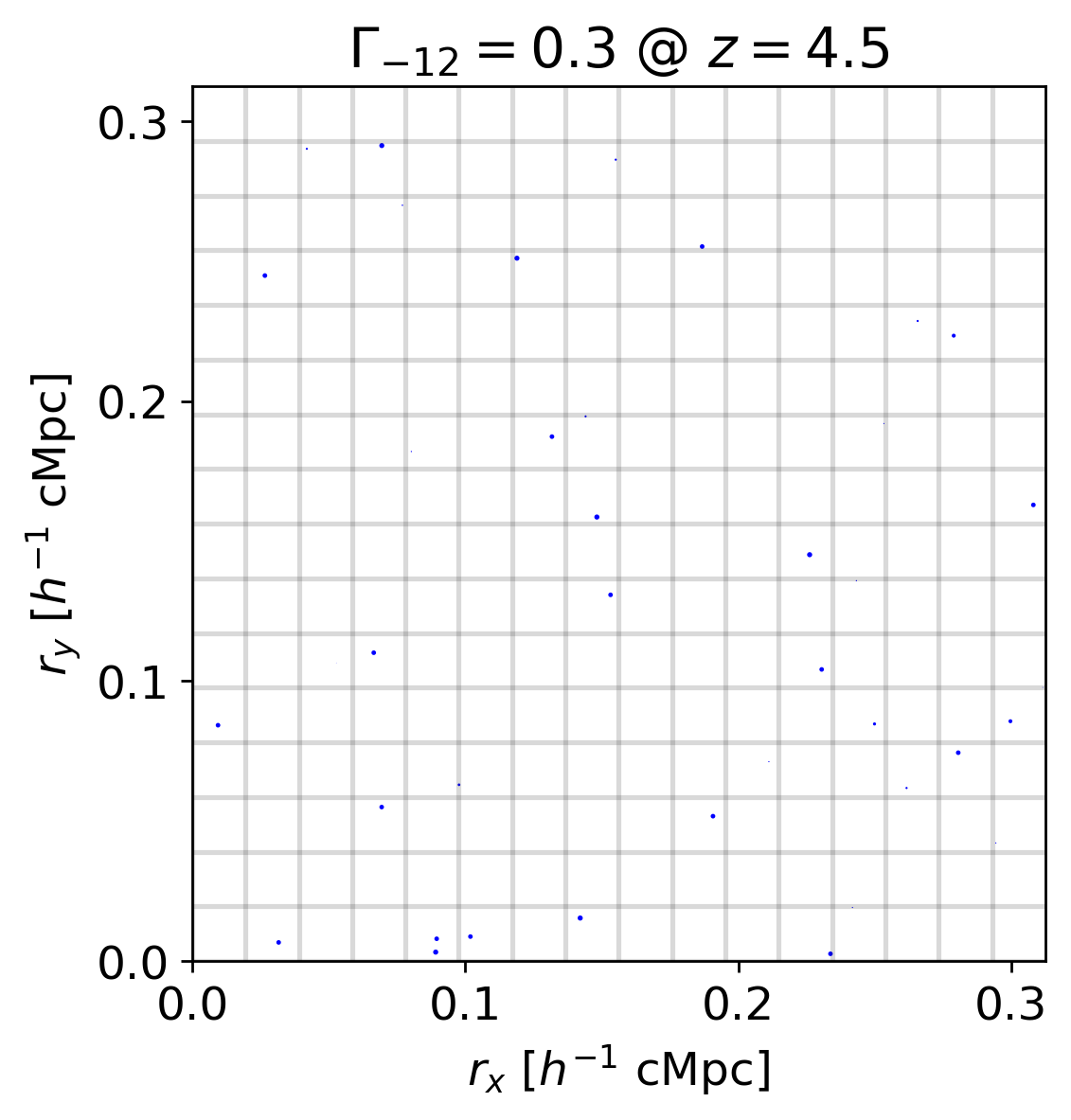}
  \caption{Projected spatial distribution of MHs in a $0.3\times0.3~[h^{-1}~{\rm cMpc}]^2$ region at $z=5.5$ (left), $5$ (middle), and $4.5$ (right) assuming $\Gamma_{-12}=0.03$ for $z=5.5$ and $\Gamma_{-12}=0.3$ for $z=5$ and 4.5. The extent of the filled circle marks the high-density core exceeding the damping-wing absorption threshold of $2\times 10^{20}~{\rm cm}^{-2}$. The grid marks the mesh of the Nyx simulation. }
   \label{fig:MHmap}
  \end{center}
\end{figure*}

First, we calculate the Ly$\alpha$ opacity in redshift space from the output HI density/velocity/temperature field of the Nyx simulation, assuming that the $z$-direction is the line-of-sight direction. However, this calculation does not include the opacity from self-shielded MHs since Nyx assumes that the cells are optically thin to the ionizing background. Therefore, we assign the HI column density to the MHs based on the 1D simulation results described in Section~\ref{sec:MH_evaporation}. 

Next, we identify MHs in the snapshots of the Nyx simulation at $z=4.5,~5,$ and $5.5$ using the Friends-of-Friends (FoF) algorithm implemented in the Nbodykit package\footnote{https://github.com/bccp/nbodykit}. We set the linking length to 0.2 times the mean particle separation. Among the FoF halos identified, we consider those with less than $5\times 10^7~M_\odot$ as MHs. We identify FoF groups down to 6 particles or $4\times10^6~M_\odot$ in mass. It should be noted that the accuracy of halo findings is not guaranteed at such small numbers of particles, but the identified particle groups give a reasonable estimate for the locations of MHs. 

We compare the mass function of the FoF halos at $z=5.5$ to the Sheth-Tormen mass function in Figure~\ref{fig:MF}. Although the two cases show a reasonable agreement, our simulation underproduces halos at around $10^7~M_\odot$ by up to a factor of 3. This mass is in the middle of MH mass range considered in this work ($4\times10^6-5\times10^7~M_\odot$). While we do not aim for a precise recovery of the Sheth-Tormen function given that the mass function (MF) is poorly constrained at such low mass and high redshift, we shall consider a possible underestimation of MH absorption due to this difference when analyzing our results.

\begin{table*}[] 
\centering
\begin{tabular}{llll}
\hline
$z$ & $5.5$ & $5$ & $4.5$  \\ \hline
No MHs             &  $0.1048$ & $0.2010$ & $0.2976$ \\ 
$\Gamma_{-12}=0.3$ &  $0.1043~(-0.47\%)$ & $\bf{0.2006~(-0.31\%)}$ & $\bf{0.2974~(-0.12\%)}$ \\
$\Gamma_{-12}=0.03$&  $\bf{0.1011~(-3.2\%)}$ & $0.1965~(-2.6\%)$ & $0.2935~(-2.1\%)$ \\ \hline
\end{tabular} 
\caption{Volume-averaged normalized Ly$\alpha$ flux for no-MH cases and MH-included cases with $\Gamma_{-12}=0.03$ and $0.3$ calculated for $z=5.5$, $5$, and $4.5$. For the MH-included cases, we also provide the fractional flux decrement with respect to the no-MH case of the corresponding redshift in parentheses next to the transmission value. We have boldfaced the results for the realistic combinations of $\Gamma_{-12}$ and $z$.}
\end{table*} \label{tab:MeanTrans}

For every identified MH, we calculate the exposure time to the ionizing radiation based on the reionization redshift field. We assume the MH had a constant mass since its exposure to the radiation. We also assume a constant UVB intensity during exposure.
Using the halo mass and exposure time of each identified MH, we calculate the HI column density profile based on the 1D simulation results described in Section~\ref{sec:MH_evaporation}. This allows us to determine the HI column densities of intervening MHs along any given sight line and for two ionizing intensities: $\Gamma_{-12}=0.03$ and $0.3$. When calculating the Ly$\alpha$ opacity due to the MHs, we treat them as point-like absorbers with a temperature of 10,000 K. This choice is made considering that Ly$\alpha$ opacity is insensitive to these parameters in the damping-wing regime. By doing so, we properly account for the absorption by MHs in the Ly$\alpha$ flux data obtained from our simulation. 

Figure~\ref{fig:MHmap} displays the projected distribution of MHs on the $xy$ plane, covering a region of $0.35\times 0.35 ~(h^{-1}~{\rm Mpc})^2$ plane along the line of sight throughout the simulation box ($80~h^{-1}~{\rm Mpc}$). The size of the circle indicates the dense core where damped Ly$\alpha$ absorption occurs (i.e., $N_{\rm HI}>2 \times 10^{20}~{\rm cm^{-2}}$). If a sight line intersects with one of these circles, the corresponding MH will appear as a DLA in the Ly$\alpha$ forest. The DLA cores have sizes of $\sim1~h^{-1}~{\rm kpc}$, which is smaller than the cell size of our simulation ($20~h^{-1}~{\rm kpc}$), described as the grid in the figure. When calculating the Ly$\alpha$ flux on the mesh data with MH absorption, we assume that all the sight lines are passing through the center of each cell.

The UVB in our Ly$\alpha$ forest simulation evolves over time, as should be the case for the real Universe (see Fig. 15 of \citealt{2017ApJ...837..106O}). Thus, the constant $\Gamma_{-12}$ assumed in our MH opacity calculation needs to be close to the time average of the evolving UVB over the photoevaporation time scale to yield realistic results.
Reionization occurs between $z=9$ and $5.5$ in our model, spanning nearly 500 Myr (Fig.~\ref{fig:Xfrac}), and photoevaporation can also take several hundred Myr when the UVB is weak ($\Gamma_{-12}\sim 0.03$) or the minihalo is on the massive end ($M_h>10^7 M_\odot$), as shown in Figure~\ref{fig:MH_evaporation}. At $z=5.5$, the time-averaged $\Gamma_{-12}$ can be much lower than the observed post-reionization value ($\Gamma_{-12}\gtrsim 0.3$), as photoevaporation of MHs may have taken place under a much weaker UVB of the reionizing Universe for most of the time, or it can also be as high as the post-reionization value if the UVB intensity does not fall steeply toward high-$z$. Thus, we mainly consider the $\Gamma_{-12}=0.03$ case at $z=5.5$, but also consider the $\Gamma_{-12}=0.3$ case  to explore the dependence of the Ly$\alpha$ forest on UVB intensity. For $z=5$ and $4.5$, $\Gamma_{-12}=0.3$ should be a realistic choice as they are well into the post-reionization era.

We also note that the Ly$\alpha$ forest is insensitive to the UVB model if the mean opacity is rescaled to a certain target value as demonstrated in \citealt{2015MNRAS.446.3697L} (see their Sec.~7). Since the mean flux will be constrained through observations in practice, the actual value of the UVB intensity used in the Ly$\alpha$ forest simulation is unimportant.

\begin{figure*}
  \begin{center}
    \includegraphics[scale=0.7]{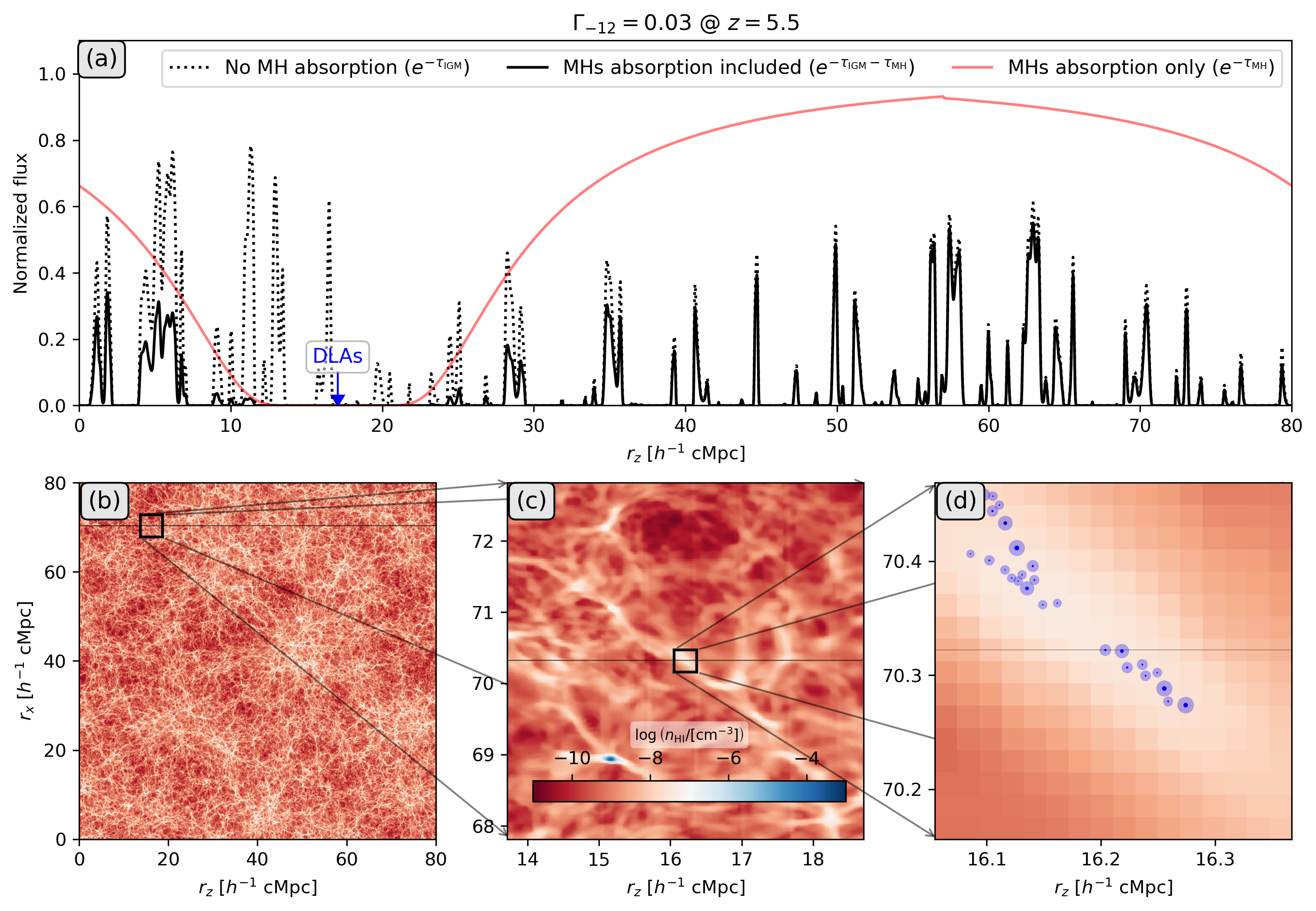}
  \caption{Panel $(a)$: Ly$\alpha$ flux of an example sight line in the $z=5.5$ snapshot of the simulated IGM. We compare the flux without MH absorption (black dotted) vs. with MH absorption assuming $\Gamma_{-12}=0.03$ (black solid). The red solid line shows the absorption by the MHs alone. The red vertical lines mark the location of MHs intersecting with the sight line. Panel $(b)$: HI density map of a slice with the sight line shown as a black straight line. Panel $(c)$: the zoom-in HI density map of the squared region in panel $(b)$. Panel $(d)$: the zoom-in HI density map of the squared region in panel $(c)$. We show the MHs on the slice as blue circles. The lighter blue circle marks the virial radius of the MHs, and the dark ones mark the DLAs. The color scale given in panel $(c)$ also applies to panels $(b)$ and $(d)$.}
   \label{fig:MHDLA_example}
  \end{center}
\end{figure*}

\section{Impact of Minihalo Absorption on Ly$\alpha$ Forest}

In Table~\ref{tab:MeanTrans}, we tabulate the volume average of the normalized Ly$\alpha$ flux from our simulation for $z=5.5$, $5$, and $4.5$ and for three cases of MH absorption: absorption not included, and absorption included with $\Gamma_{-12}=0.03$ and with $\Gamma_{-12}=0.3$. We find that MH opacity decreases the mean flux by $2-3\%$ when $\Gamma_{-12}=0.03$ and by $0.1-0.5\%$ when $\Gamma_{-12}=0.3$. The impact of MH absorption subsides toward low redshift as their HI gas photoevaporates over time. The absorption may be a factor of few stronger if we consider the underproduction of MHs in our simulation. In any case, the MH absorption effect on the mean opacity would be negligibly small for $z=5$ and $4.5$, where $\Gamma_{-12}\gtrsim 0.3$.

In Figure~\ref{fig:MHDLA_example}, we present the Ly$\alpha$ flux along an example sight line in the $z=5.5$ snapshot, demonstrating a notable absorption caused by self-shielded MHs assuming $\Gamma_{-12}=0.03$. Panels $(c)$ and $(d)$ focus on the two neighboring MHs with $8.6\times 10^6$ and $1.6\times10^7~M_\odot$ at $x\approx 16.2~h^{-1}~{\rm Mpc}$, where several MHs cluster along a filamentary structure. These two MHs exhibit $N_{\rm HI}=10.0\times 10^{20}$ and $4.1\times 10^{20}~{\rm cm}^{-2}$ for this sight line and appear as a single DLA with $N_{\rm HI}=14.1\times 10^{20}~{\rm cm}^{-2}$ in the spectrum due to their proximity. 

This example qualitatively highlights the impact of a self-shielded system on the Ly$\alpha$ forest. MHs typically form in overdense regions with $n_{\rm HI}>10^{-8}~{\rm cm}^{-3}$, where the Ly$\alpha$ flux is negligible regardless of the absorption by MHs. However, the absorption by the MHs can extend to $\sim10~h^{-1}$ Mpc, significantly changing the average flux and large-scale shape of the Ly$\alpha$ forest. Panel $(d)$ demonstrates the clustering of MHs within an overdense structure, suggesting that sight lines are more likely to encounter self-shielded systems near massive objects. Each aspect of MH absorption seen in this example will be investigated further in this section.

\subsection{DLA Incidence Rate}

\begin{figure*}
  \begin{center}
    \includegraphics[scale=0.55]{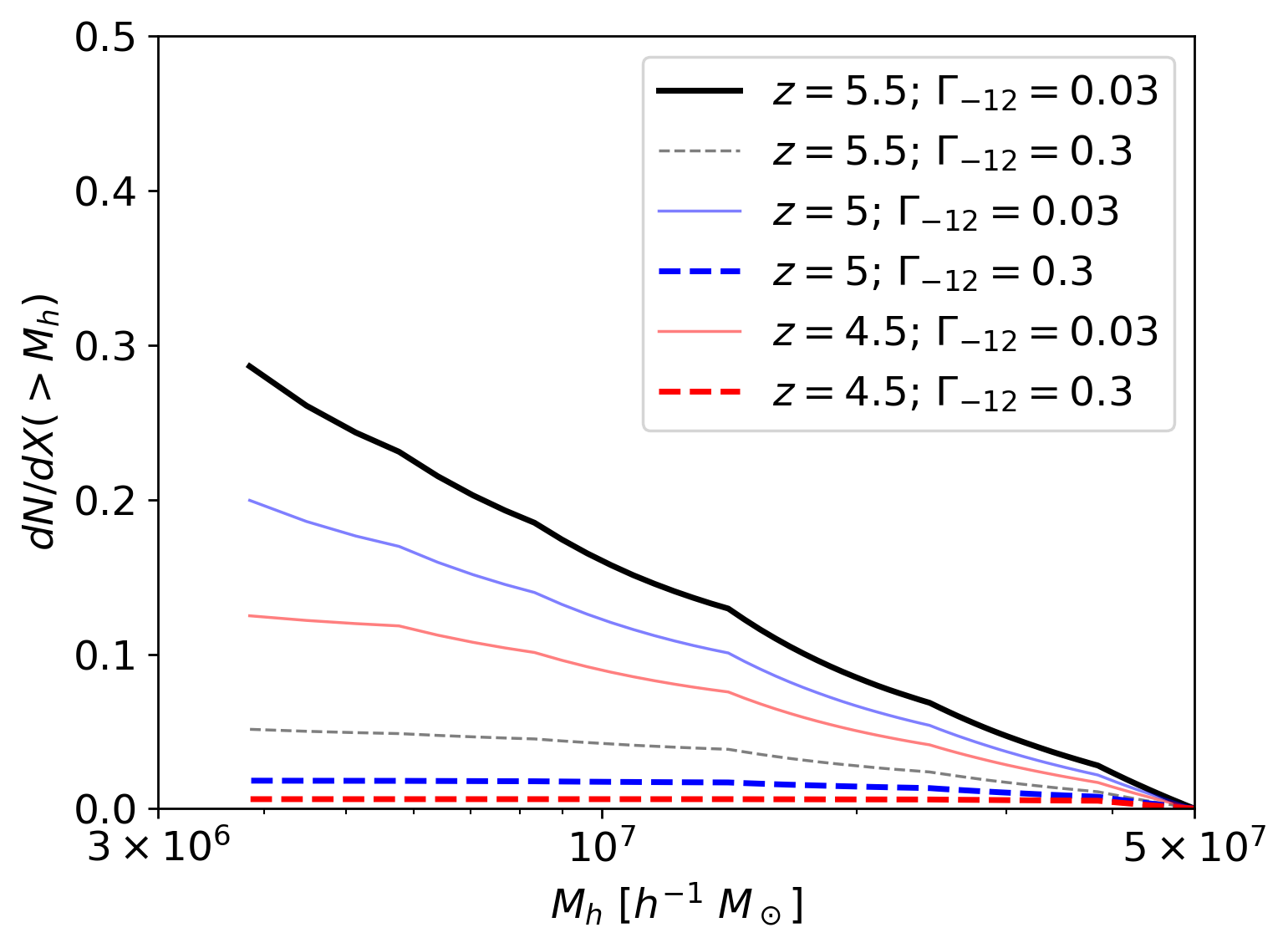}
    \includegraphics[scale=0.55]{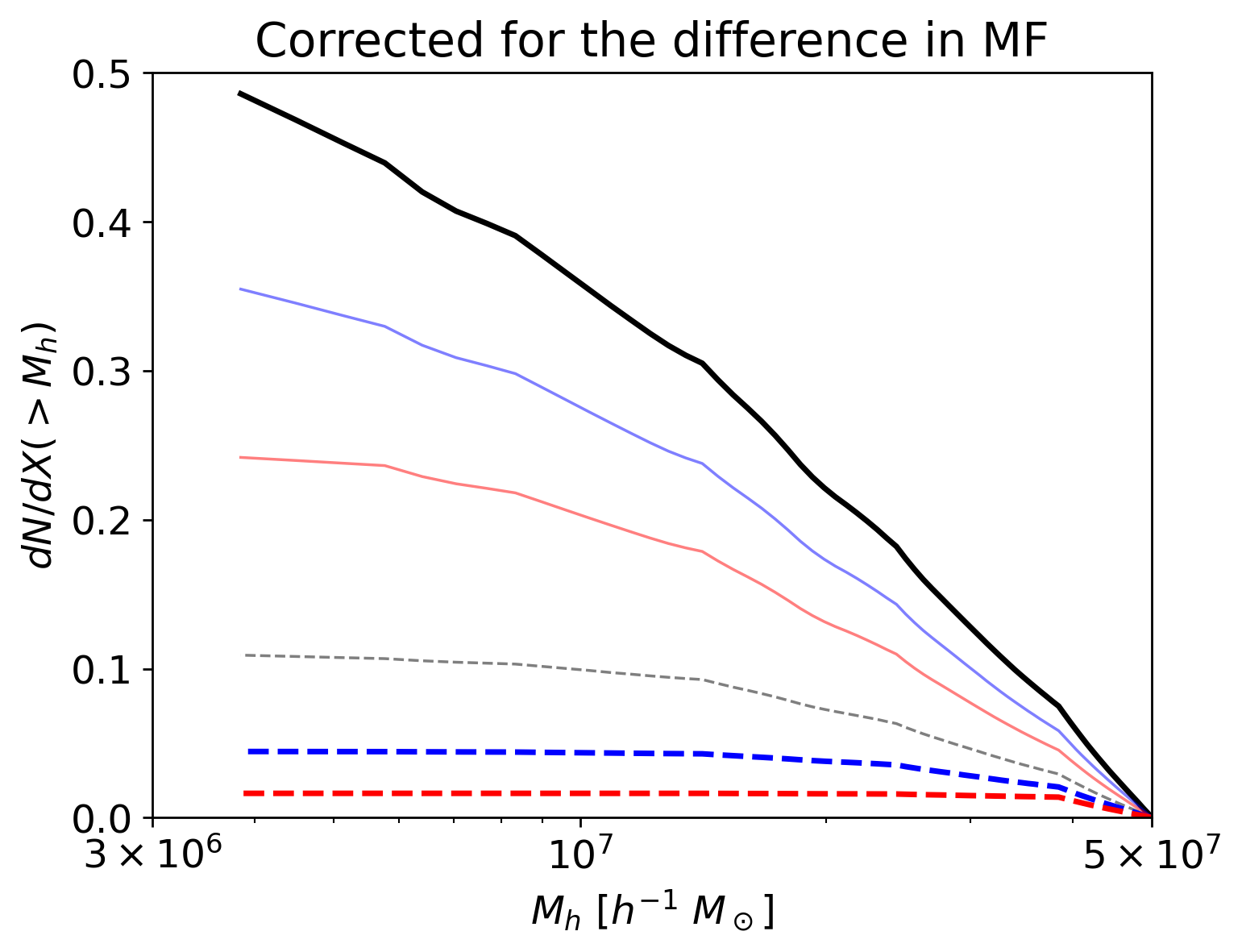}
  \caption{Incidence rate of MH DLAs per absorption distance as a function of the minimum MH mass considered. The black, blue, and red lines describe the $z=5.5$, $5$, and $4.5$ cases, respectively, and the solid and dashed lines describe the cases in which MH absorption is included assuming $\Gamma_{-12}=0.03$ and $0.3$, respectively. In the left panel, the results are directly from the simulated MHs without any correction, while in the right panel, the results are corrected for the difference between the simulated MF and the Sheth-Tormen MF.}
   \label{fig:dNdX}
  \end{center}
\end{figure*}

The latest measurement of the DLA incidence rate comes from the SDSS BOSS survey. The survey finds the incidence rate gradually increases from $dN/dX=0.033$ at $z=2$ to $0.069-0.106$ (68\% limit) at $z\sim 4.5$, where $X\equiv \int (1+z)^2 H_0/H(z) dz$ is the absorption distance \citep{2021MNRAS.507..704H}. These DLAs are considered to be associated with galactic disks hosted by atomically cooling halos in the post-reionization Universe. Since MHs are more numerous than galaxies, self-shielded MHs can significantly contribute to the DLA number density at higher redshifts toward the end of reionization. 

We calculate the DLA incidence rate contributed by self-shielded MHs by summing the projected area of their DLA portion, illustrated as blue-filled circles in Figure~\ref{fig:MHmap}. This result is shown in Figure~\ref{fig:dNdX} as a function of the minimum halo mass considered, describing the cumulative contribution of MHs toward the low-mass limit. To address the impact of the difference between the simulated MF and the Sheth-Tormen function, as shown in Figure~\ref{fig:MF}, we also calculate the incidence rate, correcting for this difference, and show the result in the right panel of Figure~\ref{fig:dNdX}.

When considering all the simulated MHs down to $4\times 10^6~h^{-1}~M_\odot$, the additional incidence rate due to MHs before correction for the difference in MF (left panel of Fig.~\ref{fig:dNdX}) is $0.3(0.05)$ at $z=5.5$ for $\Gamma_{-12}=0.03(0.3)$. The incidence rate decreases to a small value ($\lesssim 0.02$) at $z=5$ and $4.5$ for $\Gamma_{-12}=0.3$. Considering that the measurement by BOSS indicates $dN/dX = 0.1$ at most at $z=4.5$, our results suggest a steep increase in the DLA incident rate toward $z\sim5.5$ in the case of $\Gamma_{-12}=0.03$ or a mild increase in the case of $\Gamma_{-12}=0.3$. Thus, finding a rise in $dN/dX$ toward high-$z$ can give a clue for the evolution history of the UVB near the end of reionization.

The right panel of Figure~\ref{fig:dNdX} shows the same quantities after correcting for the difference in MF in Nyx versus the Sheth-Torman profile. The incidence rate increased by a factor of $\sim2$ with similar trends, making the impact of MHs even more pronounced.

The incidence rate as a function of the MH mass shows the differential contribution from different halo masses. In the cases of $z=5$ and $4.5$, the curves flatten toward the low mass, below $M_h\sim 1\times 10^7M_\odot$. The flattening indicates that the MHs below the flattening mass are fully evaporated and do not contribute to the DLA population, and therefore, $dN/dX$ has fully converged at this mass resolution. The flattening does not occur for $\Gamma_{-12}=0.03$ and $z=5.5$ (black solid line), suggesting that the photoevaporation of low-mass ($\sim 10^6~M_\odot$) MHs is still ongoing. This finding aligns with the 1D simulation result presented in Section~\ref{sec:MH_evaporation} (Fig.~\ref{fig:MH_evaporation}), which indicates that the ionizing background with $\Gamma_{-12}=0.03$ cannot completely photoevaporate MHs even after several hundred million years of exposure.

The results here show MHs can substantially contribute to the DLA population due to their large number. Considering that $dN/dX$ has not fully converged at $4\times 10^6~h^{-1}~M_\odot$, MHs below this mass may also contribute significantly to the Ly$\alpha$ opacity at $z\sim5.5$. On the contrary, the contribution of MHs appears minimal at $z\le5$ with $\Gamma_{-12}=0.3$. In this regime, neutral gas from larger halos, which we do not account for in this calculation, is likely the main constituent of DLAs. 

\subsection{Probability Distribution of Ly$\alpha$ Opacity}
 
\begin{figure}
  \begin{center}
    \includegraphics[scale=0.8]{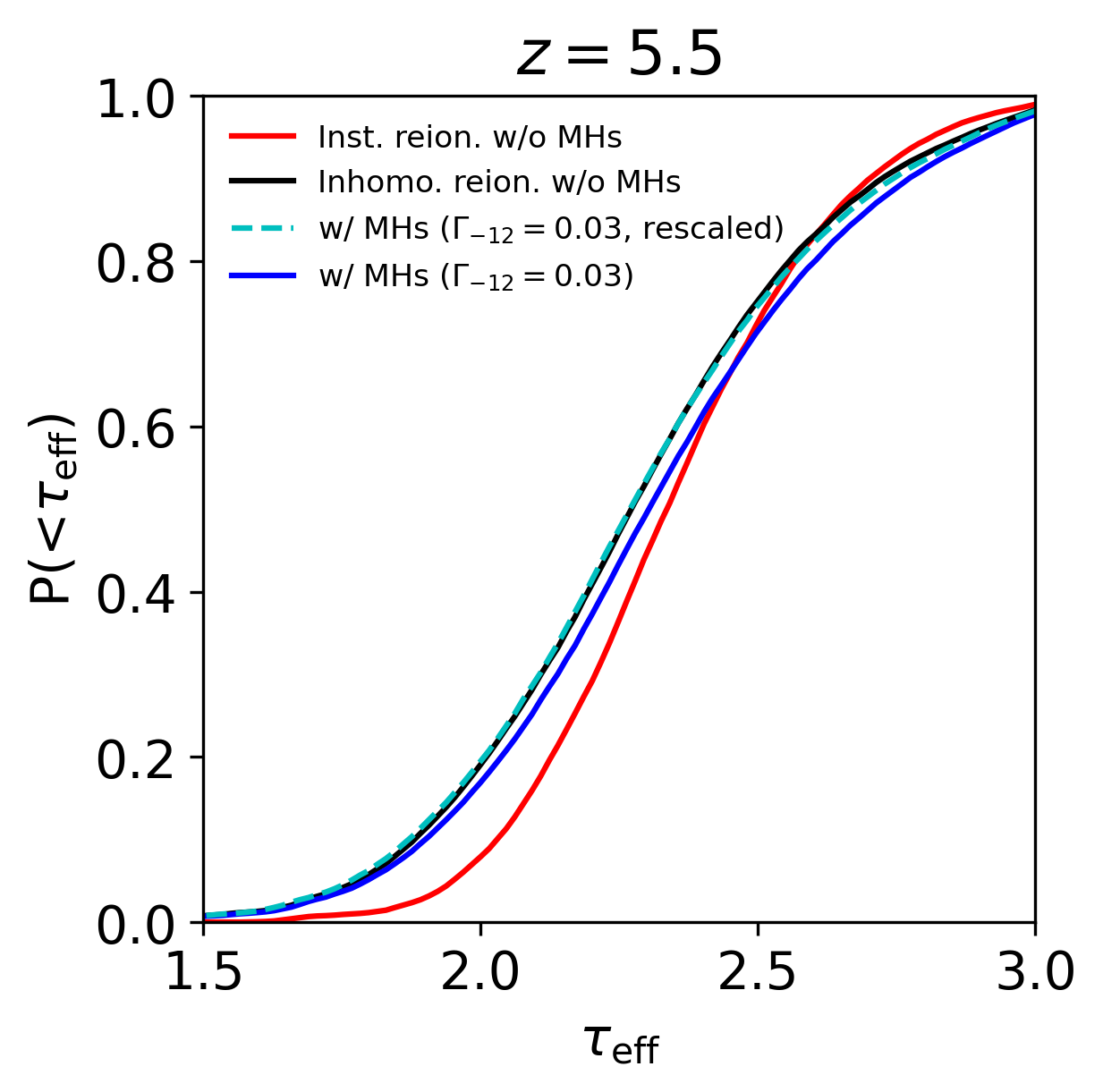}
  \caption{Cumulative probability distribution function of $\tau_{\rm eff}\equiv -\log\left<\mathcal{F}\right>$ for $z=5.5$. The black and red solid lines describe the case that we do not account for the MH absorption in the inhomogeneous and instantaneous reionization models, respectively. The blue solid line describes the case in which MH absorption is introduced to the inhomogeneous reionization model assuming $\Gamma_{-12}=0.03$. The cyan dashed line from rescaling the MH-included case to match the mean opacity with the no-MH case.}
   \label{fig:tau_eff_CDF}
  \end{center}
\end{figure}

The absorption by MHs weakens the overall Ly$\alpha$ flux by introducing extra opacity of sight lines that encounter self-shielded MHs. To quantify this effect, we calculate the effective optical depth for approximately 26,000 equally spaced skewers in our simulation. The effective optical depth is defined as $\tau_{\rm eff}\equiv-\log(\left< \mathcal{F} \right>)$, where $\left<\mathcal{F} \right>$ represents the mean flux along $50~h^{-1}~{\rm Mpc}$ line segments. This quantity is robust and unaffected by the spectral resolution, enabling a reliable comparison between different surveys or simulations. In Figure~\ref{fig:tau_eff_CDF}, we present the cumulative probability distribution (CDF) of $\tau_{\rm eff}$ with and without accounting for MH absorption. 

We compare the CDF of $\tau_{\rm eff}$ with and without accounting for MH absorption assuming $\Gamma_{-12}=0.03$ at $z=5.5$ in Figure~\ref{fig:tau_eff_CDF}. The CDF rises more slowly up to $\tau_{\rm eff}=2.5$ when MH absorption is included, indicating that a significant fraction of line segments experienced a increase in $\tau_{\rm eff}$ up to 2.5 from a lower opacity. The impact of MH absorption becomes negligibly small for $\Gamma_{-12}=0.3$. Thus, we do not show this case and lower-redshift cases ($z=5$ and $4.5$), where we expect $\Gamma_{-12}\gtrsim0.3$.

We also calculate and present the flux CDF for the instantaneous reionization case without MH absorption to assess the impact of inhomogeneous reionization in Figure~\ref{fig:tau_eff_CDF}.  These two cases exhibit agreement at the low flux limit, but the inhomogeneous reionization case converges to unity more slowly, indicating a larger number of high flux segments. This is attributed to a wider distribution of the IGM temperature in the inhomogeneous reionization case, resulting in a broader flux distribution. These findings are consistent with the findings of previous works \citep{2018MNRAS.473..560D,2022MNRAS.515.5914I}.

In practical calculations, the Ly$\alpha$ optical depth is rescaled to match the observed mean flux. To examine if the impact of MHs on CDF remains in such cases, we rescale the flux of one of the two cases (with and without MHs) to match the mean flux and assess the impact on the flux CDF. For instance, the mean flux of the case without MH absorption is $\sim3.2\%$ higher than that with the absorption for $\Gamma_{-12}=0.03$ cases (see Table~\ref{tab:MeanTrans}). We globally increase the optical depth field of the MH-included case to align the mean flux with the no-MH case. After rescaling the flux, the flux CDFs of the two cases match, suggesting the effect of MHs cannot be distinguished from the flux CDF in practice, unlike the inhomogeneity of reionization.

\subsection{Flux Power Spectrum}

At $z\gtrsim 4$, temperature fluctuations arising from inhomogeneous reionization (as illustrated in Fig.~\ref{fig:Tmap}) significantly impact the flux power spectrum at $k\sim 0.1~h^{-1}~{\rm Mpc}$, enabling to constrain the details of reionization with high-$z$ Ly$\alpha$ forest. In this regard, we evaluate the influence of MH absorption on power spectrum statistics to assess the importance of considering them in the analysis of high-$z$ quasar spectra. While at $z\lesssim4$ DLAs can be readily identified and subtracted due to their rarity and higher mean flux of the Ly$\alpha$ forest, the same is difficult at $z\gtrsim 5$ where the Ly$\alpha$ opacity is saturated in most of the cells. As a result, MH absorption features are likely to affect power spectrum analysis at those redshifts. 

\begin{figure}
  \begin{center}
    \includegraphics[scale=0.75]{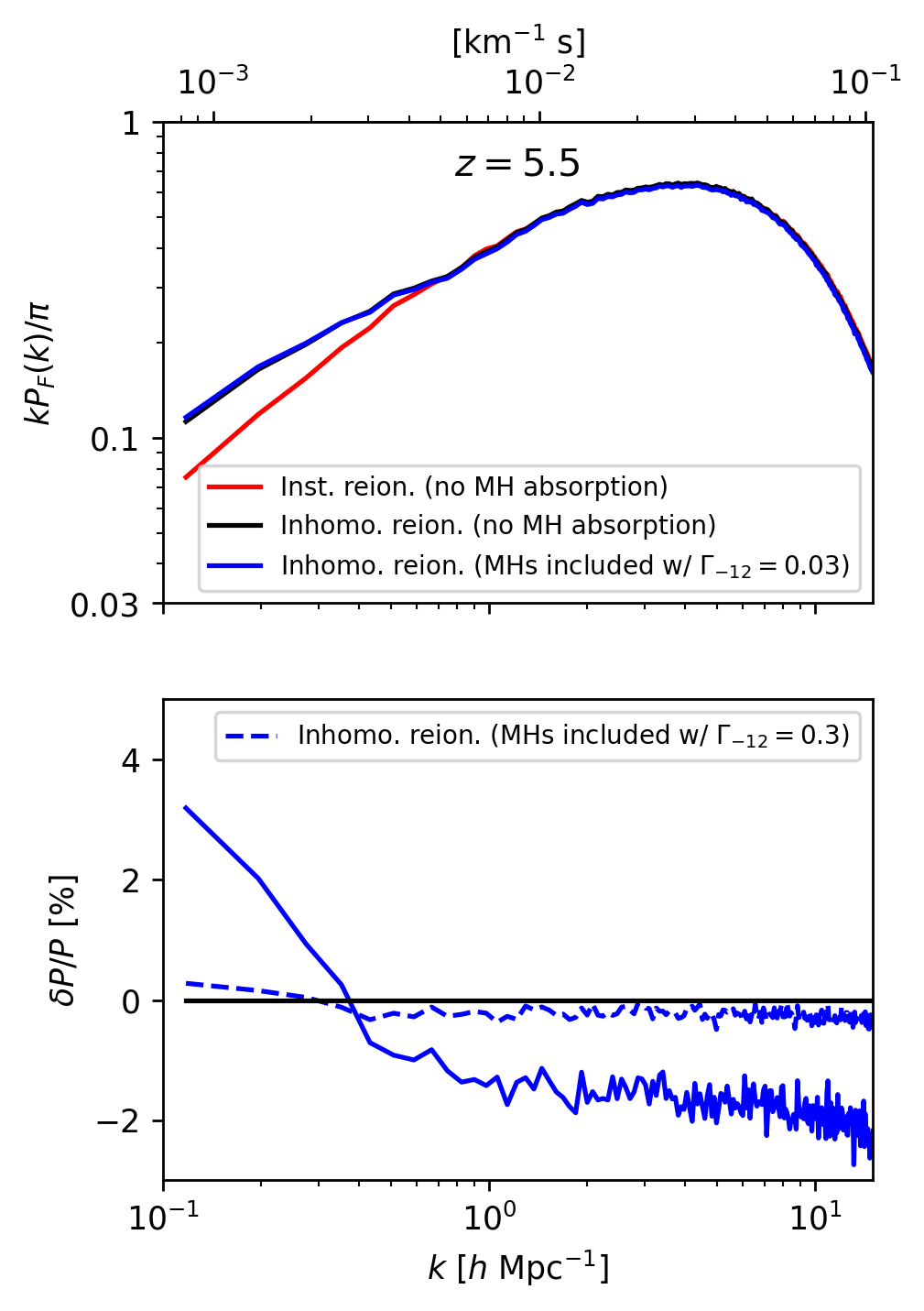}
  \caption{Upper panel: dimensionless power spectra of 1D flux, $kP(k)/\pi$, for inhomogeneous reionization run without MH absorption (black) with MH absorption assuming $\Gamma_{-12}=0.03$ (blue) and instant ($z_{\rm re} = 7$) reionization case without MH absorption (red) at $z=5.5$. Lower panel: fractional difference to the inhomogeneous reionization with no-MH absorption case. The line colors correspond to the same cases as in the upper panels, and we additionally show the result for the $\Gamma_{-12}=0.3$ case with a blue dashed line.}
   \label{fig:FluxPS}
  \end{center}
\end{figure}

We begin by rescaling the optical depth of the MH absorption-included cases to align the mean flux with that of the no-MH absorption case. Subsequently, we compute the 1D flux power spectrum for $512^2$ equally spaced skewers and present the median spectra at $z=5.5$ in Figure~\ref{fig:FluxPS}. Additionally, we plot the case of instant reionization without MH absorption and fractional difference to facilitate a comparison between the effects of inhomogeneous reionization and MH absorption. 

In the case of MH absorption with $\Gamma_{-12}=0.03$, the flux power is constantly lower than in the no-MH-absorption case by $2\%$ at $k\gtrsim 0.3~h^{-1}~{\rm Mpc}$, but it rises toward low-$k$ from $0.3~h^{-1}~{\rm Mpc}$ becoming $3\%$ higher at $0.1~h^{-1}~{\rm Mpc}$ (blue solid line in the lower panel). The extended absorption by MHs, demonstrated in Figure~\ref{fig:MHDLA_example}, adds to the large-scale power, but instead reduces the small-scale power by removing the forest in the absorbed line segment. The impact of MH absorption is negligible for $\Gamma_{-12}=0.3$ (blue dashed line), and it stays so at the other redshifts considered in this work. Thus, we do not show plots for those cases.

The wavenumber range affected by the MH absorption coincides with that affected by inhomogeneous reionization. In the comparison of the inhomogeneous reionization case (black solid) to the instant reionization case (red solid) in Figure~\ref{fig:FluxPS}, we observe that the inhomogeneity of reionization increases the flux power by 50\% at $k\sim 0.1~h^{-1}~{\rm Mpc}$ at $z=5.5$. This is fairly larger than the $-2$ to $+3\%$ modulation caused by MH absorption with $\Gamma_{-12}=0.03$ at the same redshift. 

Recalling that the MH effect on the DLA incidence rate is $\sim2$ times stronger when corrected for the underproduction of MHs in our simulation, we can speculate that the impact on the flux power spectrum could be as large as $\sim 5\%$. We also note that the instant reionization scenario is an extreme case, which likely gives a maximal difference from our inhomogeneous reionization scenario. Therefore, the MH absorption can be a subdominant but nonnegligible effect, which can introduce a bias when constraining the reionization history parameters such as the duration and midpoint with the Ly$\alpha$ forest. Given the negligible impact of MH absorption with $\Gamma_{-12}=0.3$, this complication can be avoided by utilizing the lower-redshift ($z\lesssim5$) Ly$\alpha$ forest, where the MH effect should have subsided.

\subsection{Cross correlation with Galaxies} \label{sec:corr}

In this section, we explore whether absorption by MHs can be observed in the galaxy-Ly$\alpha$ correlation signals. As the sample size of the high-$z$ galaxies and quasar sight lines continues to increase, we anticipate a significant improvement in this measurement in the coming decades.

\begin{figure*}
  \begin{center}
    \includegraphics[scale=0.7]{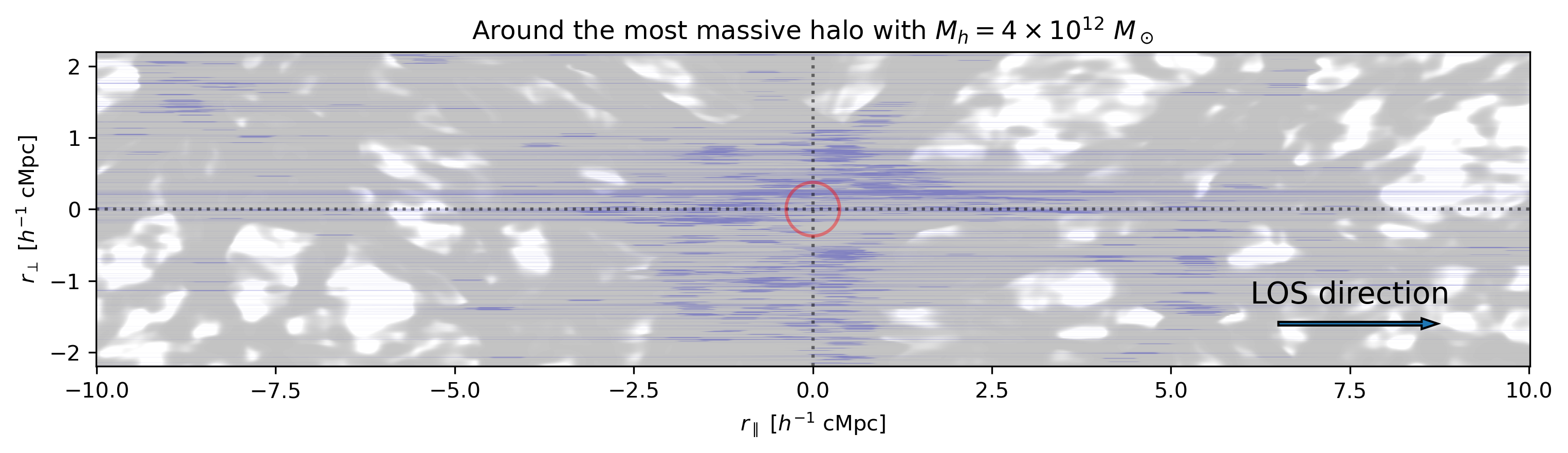}
    \includegraphics[scale=0.7]{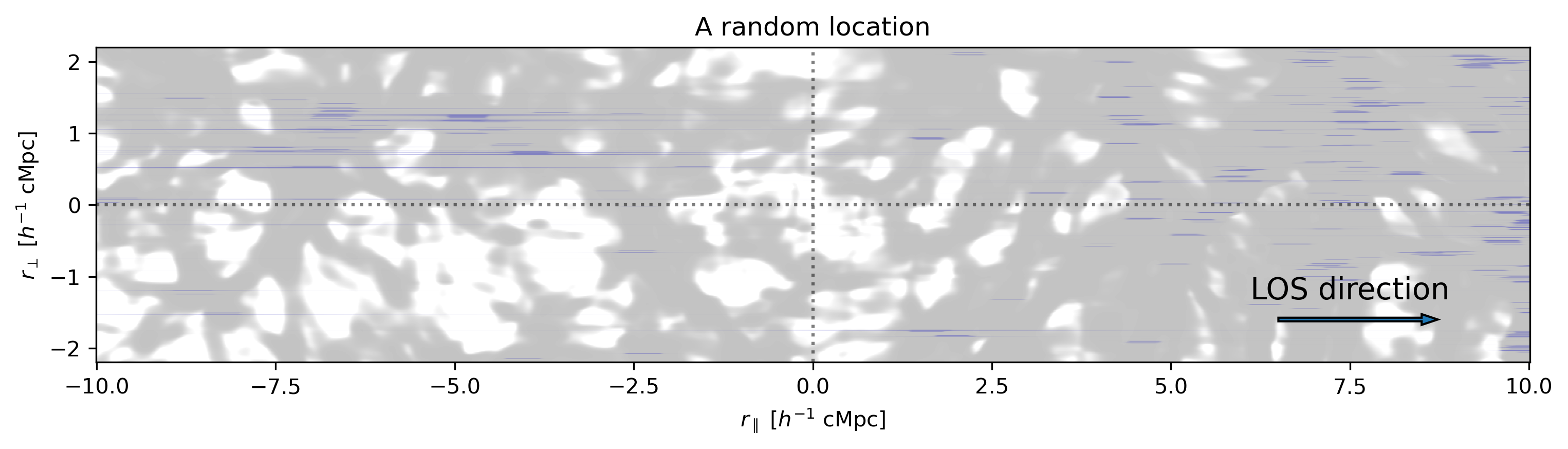}
  \caption{Ly$\alpha$ flux map around the most massive FoF halo with $M_h=4\times10^{12}~M_\odot$(upper panel) vs. a random location in the simulation (lower panel) at $z=5.5$. White regions are where the flux is significantly transmitted (30\% or above), and gray regions are where the flux is completely absorbed by the IGM without MHs. The blue stripes are where the flux is absorbed by intervening MHs in the $\Gamma_{-12}=0.03$ case. The LOS direction is along the $x$-axis, and the $y$-axis points to a direction perpendicular to the LOS. The red circle at the center of the upper panel marks the virial radius of the halo.}
   \label{fig:Flux_around_halo}
  \end{center}
\end{figure*}

As depicted in panel~$(d)$ of Figure~\ref{fig:MHDLA_example}, the MHs are highly clustered in overdense structures, implying that the absorption due to MHs would also correlate with massive structures. In Figure~\ref{fig:Flux_around_halo}, we compare the flux field around the most massive FoF halo with that at a random location along a plane parallel to the LOS direction in the $z=5.5$ snapshot for the $\Gamma_{-12}=0.03$ case. The flux map reveals a higher density of MH absorption near the halo compared to at the random location. The Ly$\alpha$ flux around the halo is already lower than the average without the MH absorption up to a few Mpc from the halo, but the extended Ly$\alpha$ absorption by DLAs extends beyond that local trend along the LOS direction up to $\sim 10~h^{-1}~{\rm Mpc}$ from the halo.

\begin{figure}
  \begin{center}
    \includegraphics[scale=0.95]{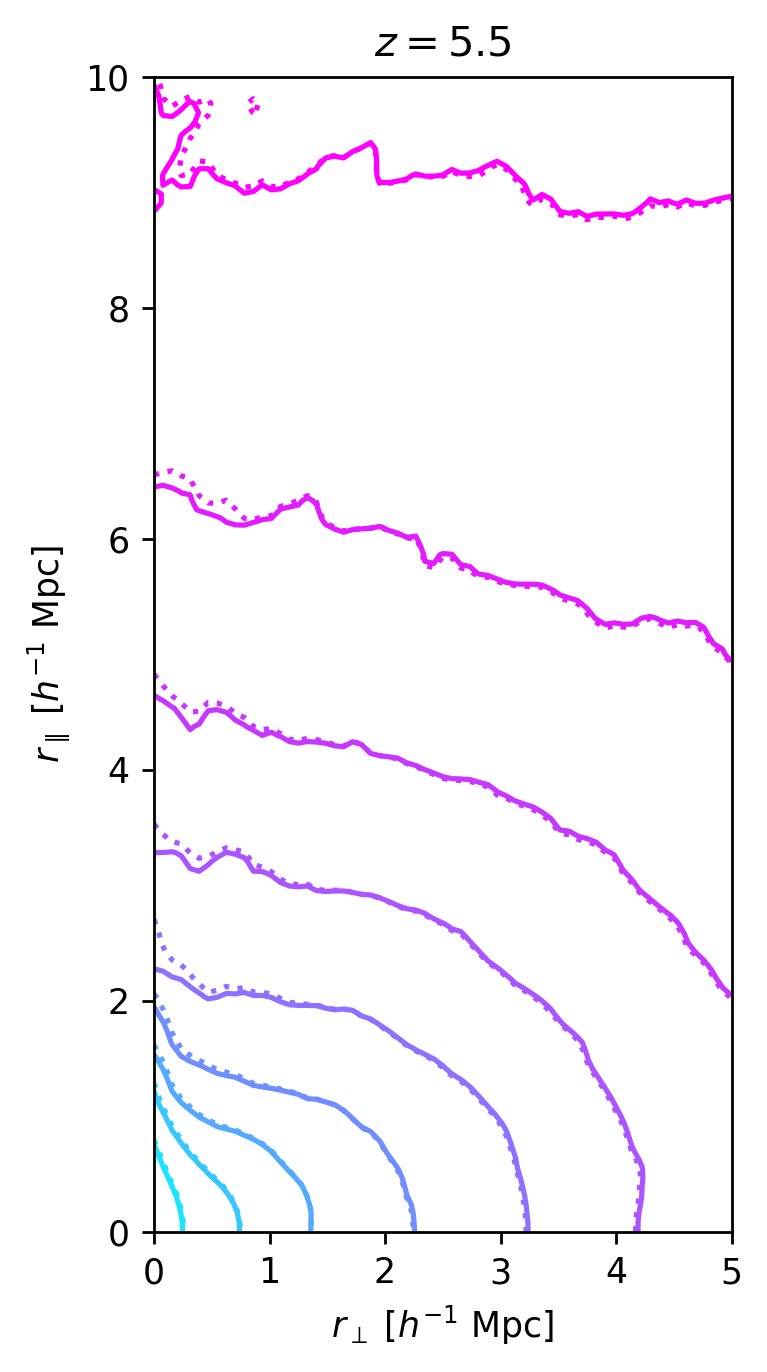}
  \caption{Stacked Ly$\alpha$ flux around 2000 most massive FoF halos shown with line contours for the $z=5.5$. The $x$ and $y$ axes are perpendicular and parallel to the LOS direction. The solid and dotted contours describe no-MH and MH-included cases with $\Gamma_{-12}=0.03$, respectively. Each case has nine line contours marking $10$, $20$, ..., $90\%$ of the mean Ly$\alpha$ flux from inside to outside. }
   \label{fig:Flux_gal_corr}
  \end{center}
\end{figure}

To quantify the spatial anticorrelation between flux and galaxies, we stack the Ly$\alpha$ flux around 2000 largest FoF halos on a grid of LOS separation ($r_\parallel\equiv\Delta r_z$) and perpendicular-to-LOS separation ($r_\perp\equiv \sqrt{\Delta r_x^2+\Delta r_y^2}$; i.e., impact parameter). The majority of the FoF halos have masses around $10^{11}~M_\odot$, corresponding to $M_{\rm UV}$ between $-19$ and $-21$. The results are shown in Figure~\ref{fig:Flux_gal_corr} for $\Gamma_{-12}=0.03$ and $z=5.5$. We do not show the result for the $\Gamma_{-12}=0.3$ case as the impact is negligible and, therefore, for lower redshifts where the ionizing background should be much stronger ($\Gamma_{-12}\gtrsim0.3$). 

The flux contours resemble the typical galaxy two-point correlation function as both galaxy and flux trace the underlying density. The contours are globally compressed along the LOS direction due to the linear gravitational motion (a.k.a. the Kaiser effect), but they are locally stretched along the LOS direction for perpendicular separation smaller than $0.5~h^{-1}~{\rm Mpc}$, extending up to $r_\parallel\sim3~h^{-1}~{\rm Mpc}$ or $\sim 300~{\rm km}~{\rm s}^{-1}$ in LOS velocity. This stretching is caused by the strong nonlinear peculiar motion of matter near the halo, also referred to as the finger-of-god (FoG) effect. 

Comparing the solid line to the dotted line illustrates the impact of MH absorption. MH absorption stretches the contours further, similar to the FoG effect, with the effects extending to larger LOS separations beyond $r_\parallel\sim3~h^{-1}~{\rm Mpc}$, reaching $10~h^{-1}~{\rm Mpc}$. 

As in the case of the DLA incidence rate, the Ly$\alpha$-galaxy correlation signal can also be $\sim2$ times stronger if we consider the underproduction of MHs in our simulation. In that scenario, the signal can be detected provided a sufficient amount of data is collected for small impact parameters ($r_\perp<1~h^{-1}{\rm Mpc}$) at high enough redshift $z\gtrsim5.5$. Additionally, the sensitivity of this cross-correlation signal to UVB intensity can be exploited to constrain $\Gamma_{-12}$ from the correlation signal.

The most relevant observation comes from \cite{2020MNRAS.494.1560M}, where the authors measured the correlation for 21 LAEs and 13 LBGs at $z\sim 6$ in the proximity of eight quasar sight lines from $z\gtrsim 6$. The data within $10$ cMpc is not sufficient to detect the signature of MHs described in Figure~\ref{fig:Flux_gal_corr}. However, they do find a significant decrease in the flux toward the galaxy at $r<10~{\rm cMpc}$, which broadly aligns with our results. With an increasing sample size, the cross-correlation signal would also be detectable in the future.

\subsection{Limitations}

Our method, as described in Section~\ref{sec:method}, unavoidably relies on several simplifying assumptions to capture subkiloparsec-scale scale physics in a volume of $\sim 100$ Mpc. While most of these assumptions are not expected to significantly alter our qualitative findings, further investigations are necessary to assess how these assumptions may affect the results. 

First, we assume the identified MHs to have the same mass since exposure to ionizing radiation. This assumption could lead to an overestimation of the MH absorption if the MH absorbing the quasar light were exposed to reionization much earlier. The time difference between the redshifts that we calculate in the Ly$\alpha$ forest ($z=5.5,~5,$ and $4.5$) and the reionization redshift in our model ($z_{\rm re}$ = $5.5 - 9$) can span several hundreds of megayears, which is longer than the typical growth time-scale for MHs \citep[e.g., see the discussion in Sec. 5.6 of][]{2020ApJ...905..151N}. In such cases, the remaining neutral gas in MHs would be overestimated. Our results in Section~\ref{sec:corr} are most likely to be affected by this assumption as massive halos are typically located in overdense regions where reionization happens earlier than in other regions. We find that the surroundings of the 2000 massive halos were ionized before $z=8$ in most cases when less than 20\% of the simulation volume was ionized. For the same reason, the calculated MHs absorption at $z=4.5$ is likely exaggerated, which would corroborate our finding that the MHs absorption is negligible at that time. 

Our calculations do not take into account any star formation inside MHs. This is considered reasonable during reionization, as the MHs are sterilized by the Lyman Werner (LW) background emitted by star-forming galaxies. However, it is possible that PopIII stars from $z\gtrsim 20$ affected the MHs during reionization via feedback. The metal enrichment resulting from the deaths of those Population III stars could enhance cooling in MHs and trigger subsequent star formation. On the other hand, MHs from $z\sim20$ might have mostly grown to atomic cooling halos by the reionization era ($z\sim 6$), or they may have evolved to non-star-forming MHs after the Population III stars ceased to exist. The quantitative impact of Population III star formation on MHs is largely unknown.

We have not accounted for the influence of the X-ray background on MHs. Hard X-rays have the ability to penetrate into the core of MHs and partially ionize the gas. This could lead to two potential effects: (a) the MH core might expand in response to increased pressure, or (b) it could contract by promoting H2 formation via the free electrons followed by H2 cooling\footnote{See also Section 5.4 of \cite{2020ApJ...905..151N} for a relevant discussion.}. A separate investigation is required to quantify this effect using methods similar to our MH evaporation simulation. 

Our simulation did not take the inhomogeneity in the ionizing background intensity into account. The background radiation is expected to be stronger at overdensities, where galaxies are clustered. The galaxy-Ly$\alpha$ correlation measurement by \cite{2020MNRAS.494.1560M} shows that the Ly$\alpha$ flux is above the mean at separations between 10 and 20 cMpc because the IGM density is close to the cosmic mean, while the ionizing intensity is above the mean at those separations. We do not find this trend in our calculation because the Ly$\alpha$ flux is calculated assuming a uniform ionizing background intensity. On the one hand, a higher mean Ly$\alpha$ flux at overdensities could lead to a stronger impact from MH absorption clustered at these locations. On the other hand, the stronger UVB background of overdense regions may result in earlier photoevaporation of MHs, thereby reducing their impact.

Lastly, as mentioned in Section~\ref{sec:method}, the constant UVB intensity assumed for the MH photoevaporation simulation is another simplification made during our calculation. The global UVB intensity is expected to rise steeply toward the end of reionization. Additionally, the intensity can locally evolve in the vicinity of massive galaxies according to their star formation rates. Therefore, the MH photoevaporation process can differ from our results, even if the fixed $\Gamma_{-12}$ matches the time average of the global UVB intensity.

\section{Summary and Conclusion}

The Ly$\alpha$ forest at $z\gtrsim 5$ is emerging as an effective probe of cosmology and reionization. During these high-redshift epochs, it is expected that the intergalactic medium contains MHs that formed prior to reionization and have yet to complete their photo-evaporation. Despite their potential significance, the impact of self-shielded MHs on the Ly$\alpha$ forest has not been systematically explored in the literature. This is partially because studying them is challenging due to their small size and the complex interplay of physical processes involved in their formation and evolution. This work addresses the gap using a hybrid scheme incorporating the HI column density of MHs, obtained from small-scale 1D simulations, into the cosmological simulation from Nyx to calculate the opacity arising from the MHs in the Ly$\alpha$ forest. We furthermore include the effect of inhomogeneous reionization in the IGM using a simple parametric model based on the ansatz from \cite{2013ApJ...776...81B} and \cite{2022ApJ...927..186T}. 

Our results are based on several simplifying assumptions that need to be tested in future studies. Ideally, one would directly resolve the MH evaporation in cosmological simulations rather than relying on those assumptions. Zoom-in or Lagrangian techniques may be required to cover the large dynamic range required for such simulations.

The impacts of MHs in our study are summarized as follows.
\begin{itemize}
    \item{The incidence rate of DLAs, $dN/dX$, increases steeply toward high-$z$ from $z\sim4.5$ to $5.5$ by $\gtrsim 0.1$.}
    \item{The Ly$\alpha$ flux is decreased by up to $\sim5\%$. }
    \item{The 1D flux power spectrum is enhanced by up to $\sim 5\%$ at $k<0.1~h~{\rm Mpc}^{-1}$ (or $10^{-3}~{\rm km}^{-1}~{\rm s}$).}
    \item{ Flux around massive halos at short perpendicular-to-LOS separation ($r_\perp \lesssim 2~h^{-1}~{\rm cMpc}$) can be particularly more suppressed due to clustered MHs in dense environments. }
    \item{These impacts are pronounced when assuming $\Gamma_{-12}=0.03$ for MH photoevaporation (possible near the end of reionization) but diminish to negligible levels for $\Gamma_{-12}=0.3$, anticipated at lower redshifts.}
\end{itemize}

In conclusion, MHs can significantly influence the statistics of the high-$z$ Ly$\alpha$ forest in the period shortly after the end of reionization. However, this effect is expected to diminish within $\sim10^8$ yr as the photoevaporation of HI gas in MHs progresses, particularly from the low-mass end, with the rapid rise of the ionizing background intensity after the end of reionization. Consequently, it is advisable to account for these effects when analyzing quasar spectra at $z\gtrsim 5.5$, whereas they are less impactful at $z\lesssim 5$. Conversely, measuring the signature of MH absorption at $z\gtrsim 5.5$ would provide insights into the ionization status of MHs and the UVB intensity during that era.

\section*{Acknowledgement}
The authors thank K. Kakiichi, N. Yoshida and the anonymous referee for the helpful comments on this work.
This work was partially supported by the DOE’s Office of Advanced Scientific Computing Research and Office of High Energy Physics through the Scientific Discovery through Advanced Computing (SciDAC) program. The development of Nyx as an AMReX application was supported by the U.S. Department of Energy, Office of Science, Office of Advanced Scientific Computing Research, Applied Mathematics program under contract number DE-AC02005CH11231, and by the Exascale Computing Project (17-SC-20-SC), a collaborative effort of the U.S. Department of Energy Office of Science and the National Nuclear Security Administration.  This research used resources of the National Energy Research Scientific Computing Center, a DOE Office of Science User Facility supported by the Office of Science of the U.S. Department of Energy under Contract No. DEC02-05CH11231. This research also used resources from the Argonne Leadership Computing Facility, which is a DOE Office of Science User Facility supported under Contract DE-AC02-06CH11357.

\bibliographystyle{apj}
\bibliography{reference}

\end{document}